\RequirePackage{fix-cm}
\documentclass[aps,twocolumn,preprintnumbers,showpacs,showkeys,nofootinbib%
]{revtex4}
\usepackage{amssymb,amsmath,amsfonts,amsthm,graphicx,psfrag}

\usepackage{graphicx}
\usepackage{epsfig}
\usepackage{color}
\usepackage{slashed}
\graphicspath{{./Figures/}}                 

\newcommand{\noi}{\noindent}
\newcommand{\beq}{\begin{equation}}
\newcommand{\eeq}{\end{equation}}
\newcommand{\bea}{\begin{eqnarray}}
\newcommand{\eea}{\end{eqnarray}}

\newcommand{\sub}[1]{_{\mbox{\small {#1}}}}

\newcommand{\ds}{\displaystyle}

\begin{document}

\title{Gluon propagators in $2+1$ lattice QCD with nonzero isospin chemical potential}

\author{V.~G.~Bornyakov}
\affiliation{NRC “Kurchatov Institute”– IHEP
Protvino, 142281, Russian Federation, \\
NRC “Kurchatov Institute”— ITEP, Moscow 117218, Russian Federation, \\
School of Biomedicine, Far Eastern Federal University,  Sukhanova 8, Vladivostok, 690950, Russian Federation}

\author{A.~A.~Nikolaev},  
\affiliation{Department of Physics, College of Science, Swansea University, Swansea SA2 8PP, United Kingdom} 

\author{R.~N.~Rogalyov}, 
\affiliation{NRC “Kurchatov Institute”– IHEP
Protvino, 142281, Russian Federation} 

\author{A.~S.~Terentev}
\affiliation{National University of Science and Technology "MISIS", Leninsky Avenue 4, 119991 Moscow, Russia}
          

\begin{abstract}
The static longitudinal and transverse  gluon propagators in the Landau gauge are studied in $2+1$  lattice QCD with nonzero isospin chemical potential $\mu_I$.
Parameterization  of the momentum dependence of the propagators is provided for all values of the chemical potential under study. We find that the longitudinal propagator is infrared suppressed at nonzero $\mu_I$ with suppression increasing with increasing $\mu_I$.  It is found, respectively, that the electric screening mass is increasing  with increasing $\mu_I$. 
Additionally, we analyze the difference between two propagators as a function of the momentum and thus compare interactions in chromoelectric and chromomagnetic sectors.
\end{abstract}
\keywords{gauge field theory, gluon propagator }
\pacs{11.15.Ha, 12.38.Gc, 12.38.Aw}


\maketitle 

\section{Introduction}
\label{section0}

To understand the physics of neutron star matter and results of heavy-ion collision experiments it is important to study the quark matter with isospin (isotopic) asymmetry. QCD phase diagram at nonzero
values of the isospin chemical potential $\mu_I$ has been studied recently quite intensively using  lattice QCD
\cite{Kogut:2002zg,Kogut:2004zg,deForcrand:2007uz,Detmold:2012wc,Cea:2012ev,Endrodi:2014lja,Brandt:2017zck,Brandt:2017oyy,Brandt:2018wkp,Braguta:2019noz,Brandt:2019hel},  
effective models as NJL model \cite{Toublan:2003tt,Barducci:2004tt,He:2005nk,He:2005sp,Ebert:2005cs,Sun:2007fc,Xia:2013caa,Khunjua:2018sro,Khunjua:2019nnv}, quark meson model \cite{Kamikado:2012bt,Wang:2017vis,Adhikari:2018cea}, chiral perturbation theory \cite{Son:2000xc,Splittorff:2000mm,Adhikari:2019zaj}, perturbative QCD \cite{Andersen:2015eoa,Graf:2015pyl}, random matrix model \cite{Klein:2003fy}, hadron resonance gas model \cite{Toublan:2004ks}
and other approaches, see e.g. recent review \cite{Mannarelli:2019hgn}.

It is also believed that the study of QCD at nonzero $\mu_I$ might help, in one way or another, to solve more difficult problem of  QCD at nonzero baryon chemical potential. 
First, QCD with nonzero $\mu_I$ is an ideal test system for some lattice approaches to QCD with nonzero baryon chemical potential such as Taylor expansion, analytical continuation and others. 
Second, lattice results obtained at nonzero $\mu_I$ by means of first principle calculations can be compared with respective results of the effective field theories of QCD and phenomenological models. Such comparison allows to test the effectiveness of QCD models and thus to estimate their applicability to the study of QCD at nonzero $\mu_B$.

The gluon propagators are among important quantities to study. Contemporary perturbative computations of the 
equation of state of high-density strong-interacting matter employ infrared behavior of gluon propagators 
described in terms of screening (see, e.g. \cite{Gorda:2021znl} and references therein).
Nonperturbative estimates of such screening effects 
on the basis of lattice simulations may be used for testing the methods of resummation of perturbation series.

In early calculations of pressure 
at nonzero quark chemical potentials 
\cite{Freedman:1976xs}, the full gluon 
propagators appear in the expressions for the 
thermodynamical potential $\Omega(T,V,\mu)$.
If the Green functions that appear in such expressions 
were evaluated nonperturbatively, the dynamics 
of pressure build-up and other thermodynamic quantities
could be investigated in detail.
 
The gluon propagator results also  will be useful in testing future results obtained at nonzero $\mu_I$ by the functional methods like Dyson-Schwinger equation and functional renormalization group.

Landau gauge gluon propagators in the non-Abelian gauge theories at zero and nonzero temperature were extensively studied in the infrared range of momenta by various methods. We shall note lattice gauge theory, Dyson-Schwinger equations, Gribov-Zwanziger approach. 
At the same time the studies at nonzero quark chemical potential are restricted to a few papers only. 
Here we close this gap for the case of nonzero isospin chemical potential.  Let us note that 
for the lattice QCD at nonzero baryon chemical potential the results can be obtained only for small values of $\mu_B$ because of the sign problem \cite{Muroya:2003qs}. 

It is known that in QCD with nonzero $\mu_I$ there is a phase transition from the normal to the superfluid phase at $\mu_I = \mu_\pi/2$.
Recently results were obtained \cite{Braguta:2019noz}  indicating that at large $\mu_I$ the theory undergoes smooth transition to the deconfinement phase. It is interesting to check if the gluon propagators change at respective values of $\mu_I$.

In this paper we make the first study of the influence of $\mu_I$ on the Landau gauge gluon propagators using lattice QCD approach. 
We use the same lattice action as in  Ref.~\cite{Braguta:2019noz}
and in fact the same set of the lattice configurations. Our goal is to study how the gluon propagators change with an increase of $\mu_I$. In particular,  we find strong suppression of the (color)-electric (longitudinal) gluon propagator with increasing $\mu_I$ which is reflected in increasing of the respective screening mass.
We parameterize the propagators in the wide range of momentum values using the fit function which is the tree level prediction of the Refined Gribov-Zwanziger approach.
We also compute an observable introduced recently in \cite{Bornyakov:2020kyz}, a difference between the (color-)electric and (color-)magnetic
propagators, and obtain its dependence on the momentum and isospin chemical potential.

The paper is organized as follows. In Section~\ref{section1} we specify details of the lattice setup to be used: lattice action, definition of the propagators and details of the simulation.
In   Section~\ref{section2}  we consider the propagators at low momenta to compute respective screening masses and determine their dependence on $\mu_I$. 
In Section ~\ref{sec:GS} the fit over a wide range of momenta is performed using Gribov-Stingl fit function motivated by the Refined Gribov-Zwanziger effective action. We demonstrate that for the longitudinal propagator this fit works well over the whole momenta domain under study for all values of $\mu_I$ considered, whereas for the transverse propagator it works only
at small $\mu_I$.

In Section~\ref{section4}  the dependence of the 
difference between the electric and magnetic
propagators on the momentum and isospin chemical potential
is discussed. We show that this difference decreases exponentially fast with momenta similarly to the case of $SU(2)$ QCD \cite{Bornyakov:2020kyz}.
The last section is devoted to the discussion of the results and to conclusions to be drawn.

\section{Simulation details}
\label{section1}

We carry out our study of the $2+1$ lattice QCD with light quark mass $m_u=m_d \equiv m_{ud}$ and strange quark mass $m_s$ using $28^4$ lattices for a set of the chemical potentials in the range $a\mu_I \in (0, 0.42)$.
The tree level improved Symanzik gauge action~\cite{Weisz:1982zw}
and the staggered fermion action with a pion source term were used to generate the lattice gauge field configurations \cite{Braguta:2019noz}.

The partition function (after integration over the quark fields) is of the form
\cite{Brandt:2017zck,Brandt:2017oyy,Brandt:2018wkp,Braguta:2019noz}
\begin{equation}
    Z = \int {DU} e^{-S_G(U)}\, det [ M^\dagger_{ud} M_{ud} + \lambda^2 ]^{1/4} det [M_s]^{1/4}\,
\end{equation}
where $\lambda$ is coupling of the pionic source term,  light quarks lattice operator $M_{ud}$ and strange quark lattice operator $M_s$ are  
\begin{equation}
M_{ud} \!= \!
 \slashed{D}(\mu_I) + m_{ud}\,, \,\,\, M_s \!= \! \slashed{D}(0) + m_{s}\,,
\label{eq:operator1}
\end{equation}
$\slashed{D}(\mu_I)$ is the staggered lattice  Dirac operator with quark chemical potential $\mu_I=\mu_u=-\mu_d$. 
The unphysical pionic source term which breaks $U(1)$ symmetry explicitly is necessary to observe the spontaneous breaking of this symmetry in a finite volume in the limit $\lambda \to 0$ at high enough $\mu_I$ \cite{Kogut:2002zg}. 
The  lattice configurations were generated \cite{Braguta:2019noz} at $\beta=4.036, am_{ud}=0.0077, am_{s}=0.0271, \lambda=0.5 am_{ud}$. 
The lattice spacing of the ensemble in physical units was determined via the Sommer scale value $r_0=0.468$~fm \cite{Bazavov:2011nk} to be $a=0.0687$~fm. Then the pion mass in physical units is $m_\pi=380$~MeV \cite{Braguta:2019noz}.

Our lattice spacing is small enough to be sure that usual
lattice discretization effects are small. However, 
at nonzero isospin chemical potential 
new lattice artifacts may appear.
To reach large quark densities without lattice
artifacts one needs sufficiently 
small lattice spacing to satisfy condition
$a\mu_I \ll 1$. Actual values of $a\mu_I$
under consideration are given in Table~\ref{tab:GSL_param_DOS18} of Appendix A,  one can see that $a\mu_I < 0.42$.
Thus we can hope that respective lattice artifacts are small.

At the same time to study 
the gluon propagators in the infrared region 
it is necessary to employ large physical volume. 
As a result of a compromise between these 
two requirements our lattice size is rather
moderate: $L = 28a = 1.92$~fm.   
This implies a potential problem of large
finite volume effects at small momenta
(the minimal momentum $p_{min}\approx 650$~MeV). 

Our lattice with equal number of temporal and spatial sites $N_s=N_t=28$ corresponds to the temperature 
$\ds T={1\over 28 a}\approx 100$~MeV. However, the energy
of the lightest excitation at $\mu_I=0$ is estimated as $E_{exc}\geq p_{min}=650$~MeV. Since $E_{exc}\gg T$, this temperature can be considered as approximately zero. 
At the greatest isospin chemical potential under study,
$\mu_I=1.2$~GeV, the energy of the lightest excitation 
can be estimated as follows. Using the model of free quarks,
we notice that in the respective ground state all modes with 
$|\vec p|<\mu_I$ are occupied by $u\bar d$ quark states 
(27 modes with $|\vec p| \leq p_{min}\sqrt{3}$ in the case under consideration). The lightest excitation occurs when one 
light quark jumps from the mode with $|\vec p|= p_{min}\sqrt{3}$ to the mode with $|\vec p|= 2p_{min}$.
The energy needed for this purpose is $E_{exc}\sim 170$~MeV, therefore, in a rough approximation, the temperature $T\sim 100$~MeV can be thought of as zero even at $\mu_I=1.2$~GeV. 

In our study of the gluon propagators we 
employ the standard definition of 
the lattice gauge vector
potential $A_{x,\mu}$ \cite{Mandula:1987rh}:
\beq
A_{x,\mu} = \frac{1}{2iag}~\Bigl( U_{x\mu}-U_{x\mu}^{\dagger}\Bigr)\bigg|_{traceless}.
\label{eq:a_field}
\eeq
The lattice Landau gauge fixing condition is
\beq
(\nabla^B A)_{x} \equiv {1\over a} \sum_{\mu=1}^4 \left( A_{x,\mu}
- A_{x-a\hat{\mu},\mu} \right)  = 0 \; ,
\label{eq:diff_gaugecondition}
\eeq

\noi which is equivalent to finding an extremum of the gauge functional

\beq
F_U(\omega)\; =\; \frac{1}{4V} \sum_{x,\mu}\ \frac{1}{3} \mathrm{Re \, Tr}\; U^{\omega}_{x\mu} \;,
\label{eq:gaugefunctional}
\eeq

\noi with respect to gauge transformations $\omega_x~$.
To fix the Landau gauge we use the simulated annealing (SA) algorithm
with finalizing overrelaxation \cite{Bogolubsky:2007bw,Bornyakov:2009ug}.

The gluon propagator $D_{\mu\nu}^{ab}(p)$ is defined
as follows: 
\beq
D_{\mu\nu}^{ab}(p) = \frac{T}{V}
    \langle \widetilde{A}_{\mu}^a(q) \widetilde{A}_{\nu}^b(-q) \rangle\;,
\qquad
\eeq
where $V=(N_s a)^3$, $\ds T={1\over N_t a}$,

\bea\label{eq:gluonpropagator}
\widetilde A_\mu^b(q) = a^4 \sum_{x} A_{x,\mu}^b
\exp  && \left( {2\pi i q_4\over N_t}\left({x_4\over a}+{\delta_{\mu 4}\over 2}\right) \right. \\ \nonumber 
&& + \left. {2\pi i q_k\over N_s}\left({x_k \over a}+{\delta_{\mu  k}\over 2}\right) \right), \nonumber
\eea
$q_i \in (-N_s/2,N_s/2]$, $ q_4 \in (-N_t/2,N_t/2]$  and
the physical momenta $p_\mu$ are defined by the relations $ap_{i}=2 \sin{(\pi q_i/N_s)}$,
$ap_{4}=2\sin{(\pi q_4/N_t)}$;  $p=\sqrt{\sum_{\mu=1}^4 p_\mu^2}$.

At nonzero $\mu_I$ the $O(4)$ symmetry is broken and
there are two tensor structures for the gluon propagator \cite{Kapusta:2006pm}~:
\beq
D_{\mu\nu}^{ab}(p)=\delta_{ab} \left( P^T_{\mu\nu}(p)D_{T}(p) +
P^L_{\mu\nu}(p)D_{L}(p)\right)\,,
\eeq

\noi where (symmetric) orthogonal projectors $P^{T;L}_{\mu\nu}(p)$
are defined for $p=(\vec{p}\ne 0;~p_4=0)$ as follows

\bea
P^T_{ij}(p)&=&\left(\delta_{ij} - \frac{p_i p_j}{\vec{p}^2} \right),\,
~~~P^T_{\mu 4}(p)=0~;
\\
P^L_{44}(p) &=& 1~;~~P^L_{\mu i}(p) = 0 \,.
\eea

\noi Therefore, two scalar propagators - longitudinal $D_{L}(p)$ and
transverse $D_T(p)$ -  are given by
\bea\label{gluonpropagator}
D_T(p)&=&\frac{1}{16} \sum_{a=1}^{8}\sum_{i=1}^{3}D_{ii}^{aa}(p)~;
 \\
D_L(p)&=& \frac{1}{8}\sum_{a=1}^{8} D_{44}^{aa}(p) \,. 
\eea
For $\vec{p} = 0$ they are defined as follows:
\bea\label{eq:DL_DT_zeromom}
D_T(0) &=& \frac{1}{24} \sum_{a=1}^{8} \sum_{i=1}^{3} D^{aa}_{ii}(0)\,,
\\
D_L(0) &=& \frac{1}{8}\sum_{a=1}^{8} D^{aa}_{00}(0)\  .
\eea
$D_T(p)$  is associated with the magnetic
sector, $D_L(p)$ -- with the electric sector.
We consider the soft modes $p_4=0$ and use the notation $D_{L,T}(p)=D_{L,T}(0,|\vec{p}|)$.

\section{Screening masses}
\label{section2}

\begin{figure*}[tbh]
\hspace*{-3mm}\includegraphics[scale=0.4]{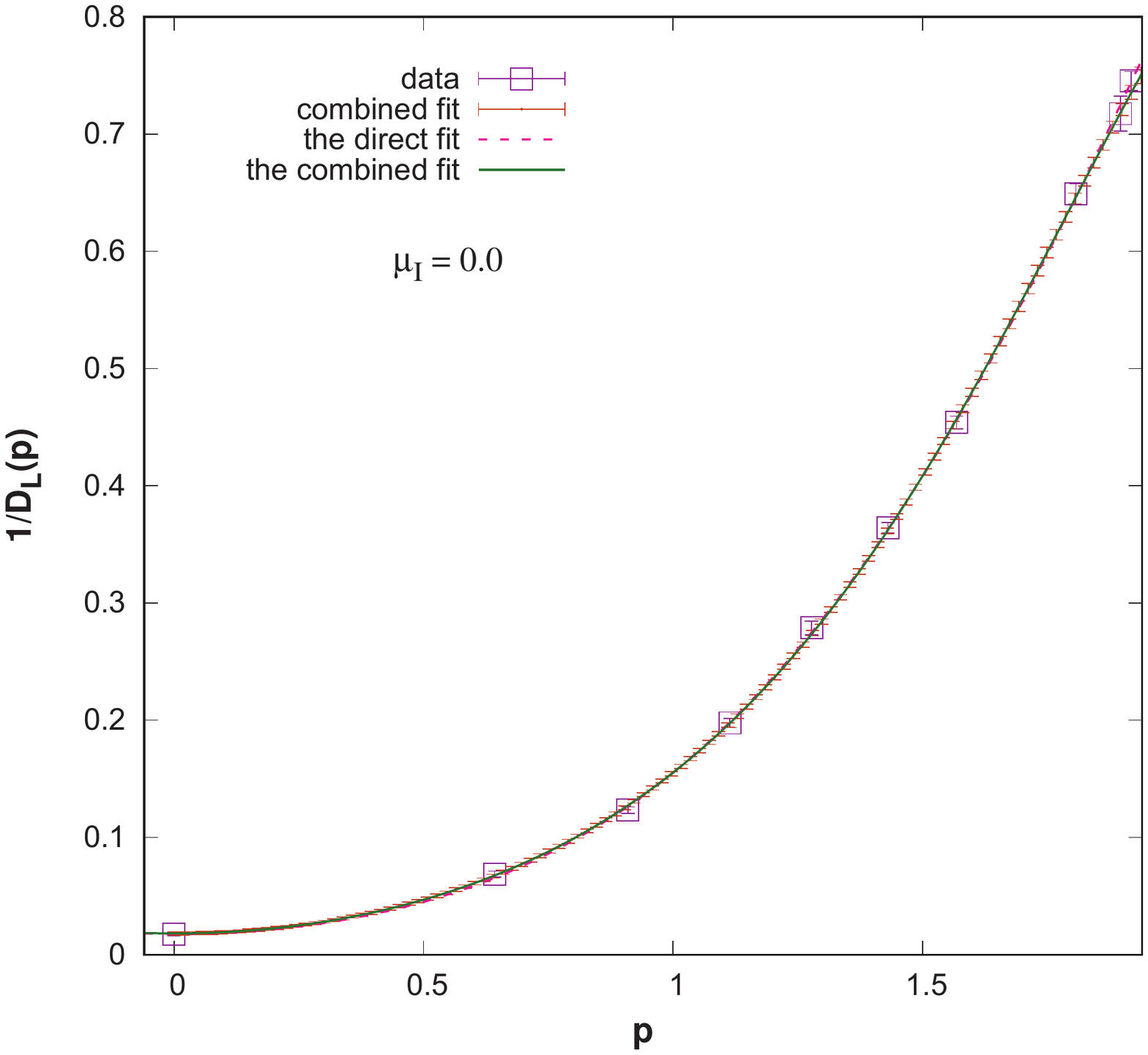} \includegraphics[scale=0.4]{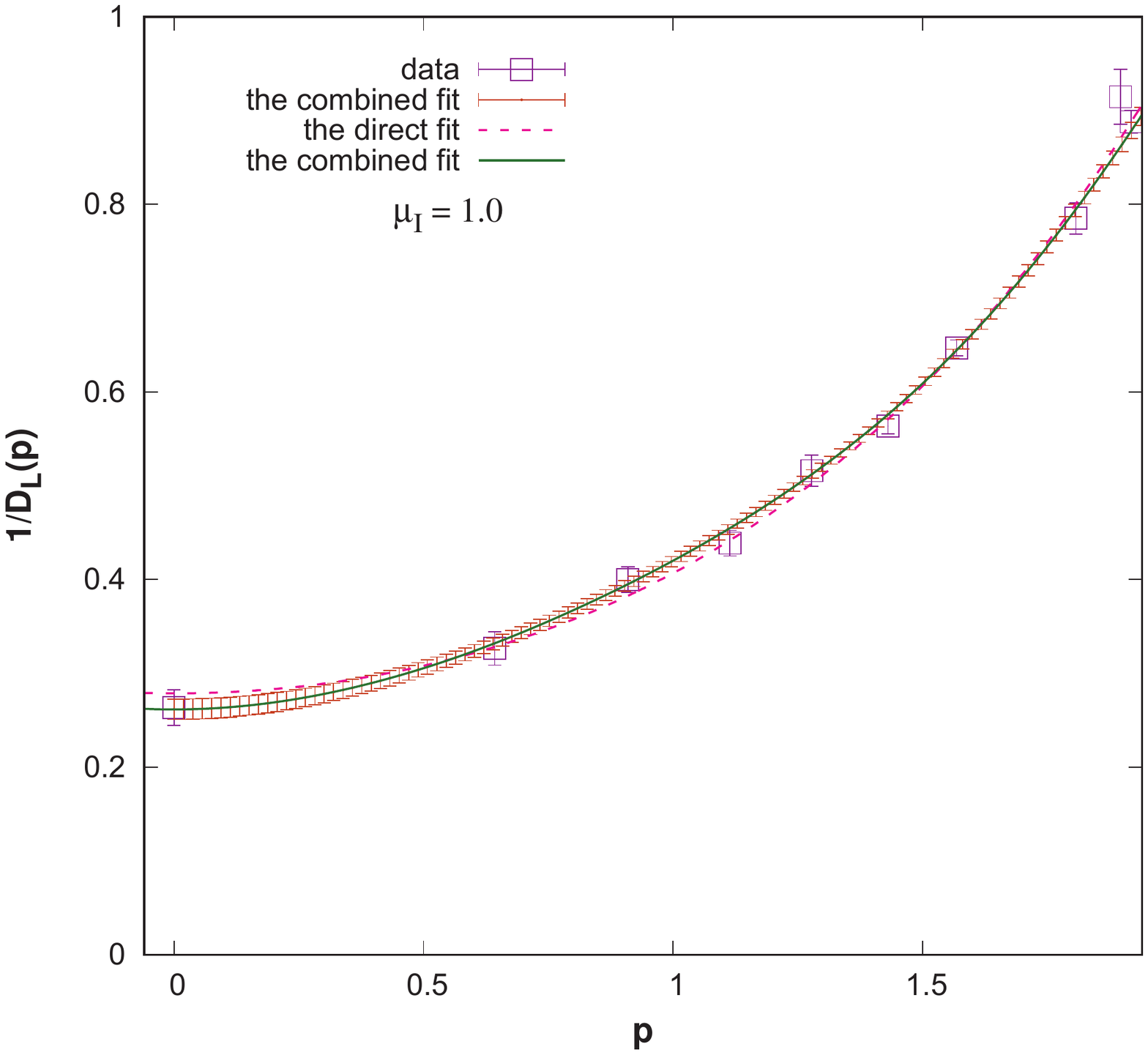}
\vspace*{-28mm}
\caption{The inverse longitudinal propagator
interpolated by formula (\ref{eq:fit_B_for_mE})  (the combined fit) or by (\ref{eq:LOW_MOM_fit_fun}) 
(the direct fit) 
at $\mu_I=0$ (left panel) and $\mu_I=1$~GeV (right panel). 
The corridors of errors in the former case are shown by the respective error bars.  Data are shown by purple squares.}
\label{fig:DLinverse}
\end{figure*}

\begin{figure*}[tbh]
\hspace*{-20mm}\includegraphics[scale=0.38]{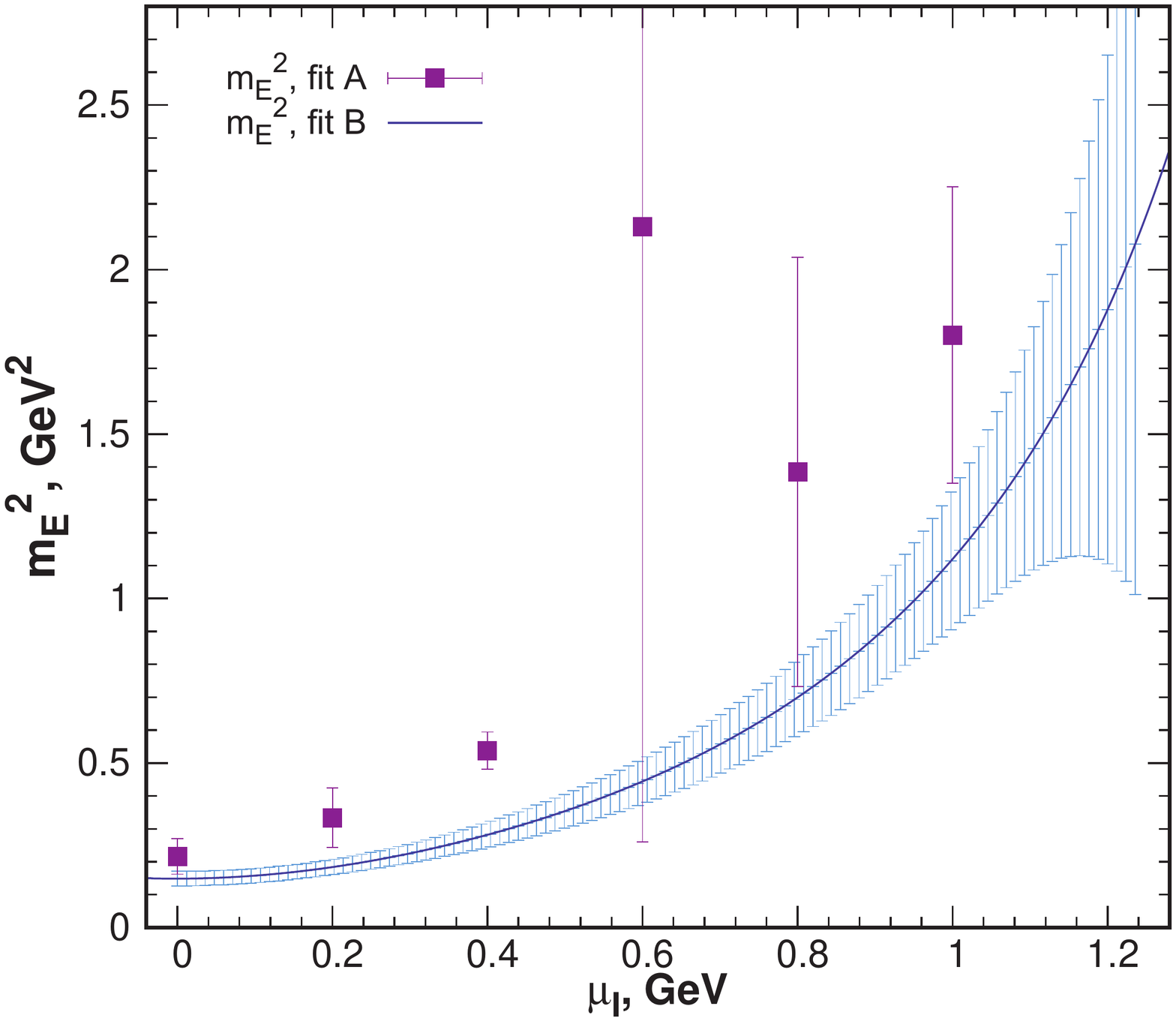} \hspace*{-19mm}
\includegraphics[scale=0.38]{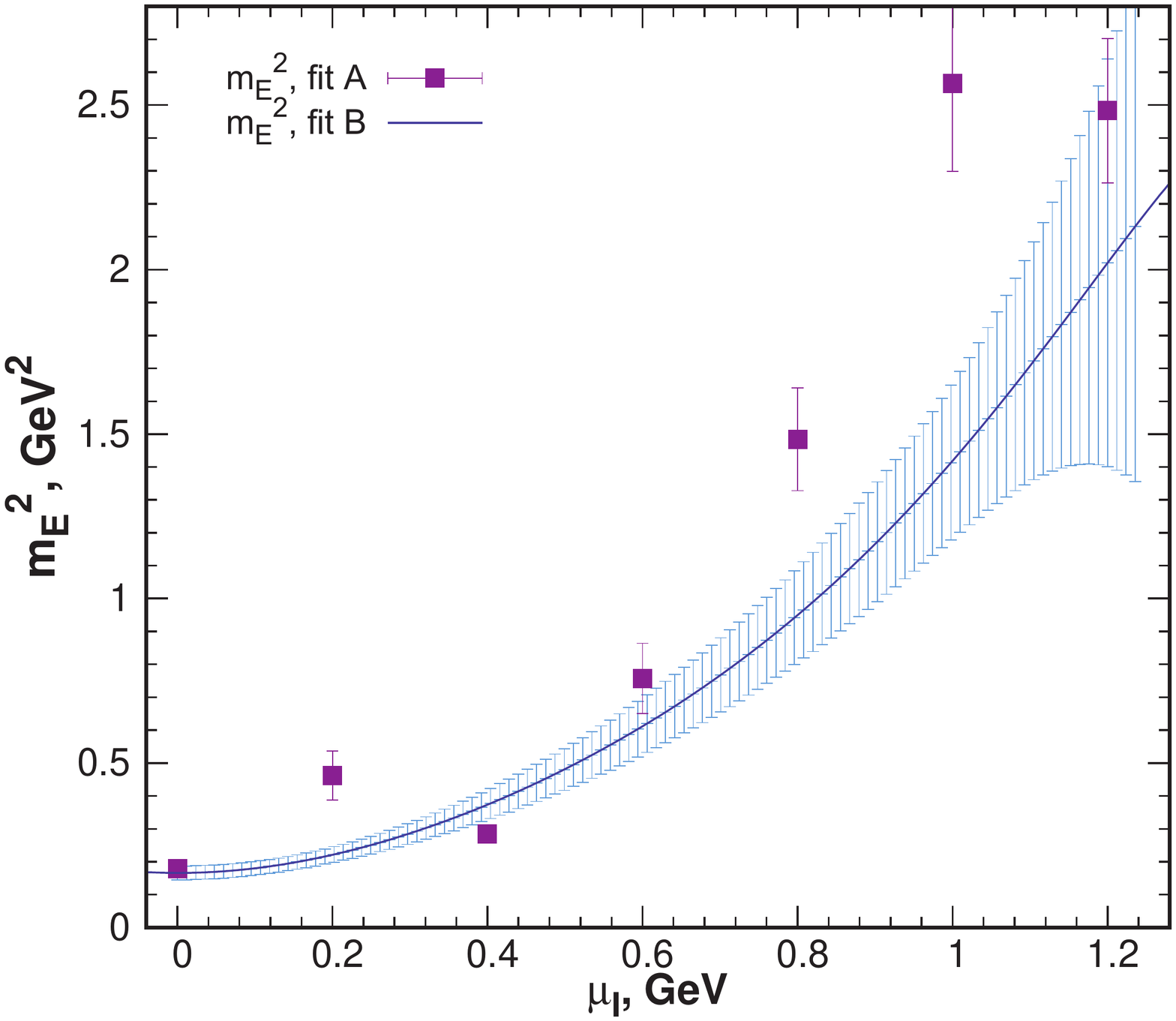}
\vspace*{-8mm}
\caption{The chromoelectric squared screening masses as a function of the isospin chemical potential. Fit A is related to the fit functions (\ref{eq:LOW_MOM_fit_fun}); 
fit B---to 
(\ref{eq:fit_B_for_mE}); the momentum ranges  are $0\leq p < 1.5$~GeV (left) and $0\leq p < 2$~GeV (right). Curves show interpolated values of the screening masses evaluated by the formula (\ref{eq:mE_interpolated}), the respective corridors of errors are shown by error bars.  }
\label{fig:mesq}
\end{figure*}

Various definitions of the screening masses were discussed in great detail in our previous studies \cite{Bornyakov:2019jfz,Bornyakov:2020kyz}.
Here we use the definition in terms of the Taylor expansion of $D_{L,T}^{-1}(p)$ in $p^2$ at low momenta:
\beq\label{eq:LOW_MOM_fit_fun}
D_{L,T}^{-1}(p) = Z^{-1}\large( \tilde{m}_{E,M}^2 + p^2 + c_4 \cdot (p^2)^2 + c_6 (p^2)^3 ... \large)\; .
\eeq
This method was used in \cite{Bornyakov:2011jm} 
in the studies of lattice QCD at finite temperature 
and was applied to QC$_2$D 
in \cite{Bornyakov:2019jfz}. 
One can note that the Yukawa-type fit-function
\beq\label{scrmass_mom1}
D_{L,T}^{-1}(p) = Z^{-1}(\tilde{m}_{E,M}^2 + p^2)
\eeq
successfully employed in the previous studies of lattice gluodynamics at zero and finite temperatures in  Refs.\cite{Bornyakov:2010nc,Oliveira:2010xc,Silva:2013maa} 
does not work in the case under consideration because 
the minimal nonzero momentum on our lattices 
is rather large: $p\sub{min}\approx 640$~MeV. 
Thus we have no data points in deep infrared domain, 
where higher-order terms in 
formula~(\ref{eq:LOW_MOM_fit_fun}) can be neglected.
As explained below, we have to take into account 
up to the $\underline{O}\Big((p^2)^3\Big)$ terms 
of the Taylor expansion of the inverse propagators. 

It should be noted that the screening masses 
$\tilde{m}_{E,M}^2$ are related to the correlation lengths:
\beq
\label{mass_def1}
\tilde{m}_{E,M}^2=\xi_{E,M}^{-2},
\eeq
where $\xi_{E,M}$ are usually defined in terms of  propagators as follows \cite{ShangKengMa}:
\bea
\label{eq:correlation}
  \xi^2 &=& \frac{1}{2} {\int_V dx_4 d\vec x  D(x_4, \vec x) |\vec x|^2
\over  \int_V dx_4 d\vec x   D(x_4, \vec x)} \\ \nonumber
&=&
- {1\over 2 D(0, \vec 0)}\; \sum_{i=1}^3
\left({d\over dp_i}\right)^2\Big|_{\vec p=0} D(0, \vec p)\; . \nonumber 
\eea

To extract the screening masses from the data, we employ
the fit function (\ref{eq:LOW_MOM_fit_fun})
keeping terms up to $p^2$ or $(p^2)^2$ or $(p^2)^3$,  depending on the fit range. We will call below respective fit functions as fit function with two terms, with three terms or with four terms.
We perform the fit over the range $p_{\mbox{low}} \leq p < p_{\mbox{cut}}$, 
where $p_{\mbox{low}}$ is equal to either zero or $p_{\mbox{min}}$
and $p_{\mbox{cut}}$ varies from 1.3~GeV to 2~GeV. 

We find that, in the general case, both $m_E$ and $m_M$ 
depend substantially on the choice of $p_{\mbox{low}}$, $p_{\mbox{cut}}$ and the fit function. To improve stability of the results,
we consider the propagators as functions of 2
variables, $D^{-1}_{L,T}(p,\mu_I)$. In so doing,
we assume smooth dependence of a propagator
on both variables and fit the overall set of data 
by a polynomial in $p^2$ and $\mu_I^2$ thus reducing drastically
the total number of estimated parameters. 
We choose the polynomial by selecting a finite number of terms in the series
\beq\label{eq:general_pow_ser}
D^{-1}_{L,T}(p,\mu_I) \simeq \sum_{n=0}^\infty \sum_{k=0}^\infty c_{2n,2k} (p^2)^n \mu_I^{2k} \;, 
\eeq
we omit the superscript $L$ or $T$ of the coefficients $c_{i,j}$. 
Odd powers of $\mu_I$ are prohibited by isospin symmetry 
of strong interactions and, in particular, 
gluon propagators are invariant under the change 
$\mu_I\;\to\;-\;\mu_I\;$.
Such polynomials providing a reliable quality of the fit
are found by trial-and-error method in the longitudinal 
case, the resulting expressions for each fit range 
are given below. 
The results of this (that is, combined) and the direct 
(that is, by formulas like (\ref{eq:LOW_MOM_fit_fun})) fits 
for the inverse propagator at a given value of $\mu_I$ are compared in Fig.~\ref{fig:DLinverse}.
It should be remembered that an increase of the number of 
estimated parameters gives rise to increasing of uncertainties in their values, 
whereas a decrease of this number 
results in either a dramatic decrease of the fit $p$-value 
or in a substantial shrinking of the fit range.

In the case of transverse propagator both fitting procedures
are not stable with respect to elimination of the zero momentum;
data on larger lattices and increased statistics are needed
to arrive at reliable results. In this study, screening 
in the magnetic sector is discussed in terms of the dressing function, see Section~\ref{section4}.

\subsection{Chromoelectric screening mass}

We fit our data by the function (\ref{eq:LOW_MOM_fit_fun}) 
with three or four terms using $p_{\mbox{cut}}=1.3, 1.5$ or $2.0$~GeV and $p_{\mbox{low}}=0$ or $p_{\mbox{min}}$.
The results of the fits are stable with respect to exclusion of the zero momentum only for $p_{\mbox{cut}}=2$~GeV.
In this case, the fit function (\ref{eq:LOW_MOM_fit_fun}) with three  terms works well at $\mu_I\geq 0.4$~GeV; at $0\leq \mu_I\leq 0.2$~GeV we employ the fit function (\ref{eq:LOW_MOM_fit_fun}) with four terms. These results are shown in the right panel of Fig.~\ref{fig:mesq} by squares. In performing this fitting procedure we determine in total 23 fit parameters (when all values of $\mu_I$ are taken into account). 

As we explained above one can substantially reduce the number of fit parameters
by considering the inverse propagator as function of $p$
and $\mu_I$. This approach also makes it possible to
derive $m_E^2$ as function of $\mu_I$ and estimate
respective confidence interval for each value of $\mu_I\in (0, 1.2$~GeV).

\begin{figure*}[hbt]
\includegraphics[width=0.48\textwidth]{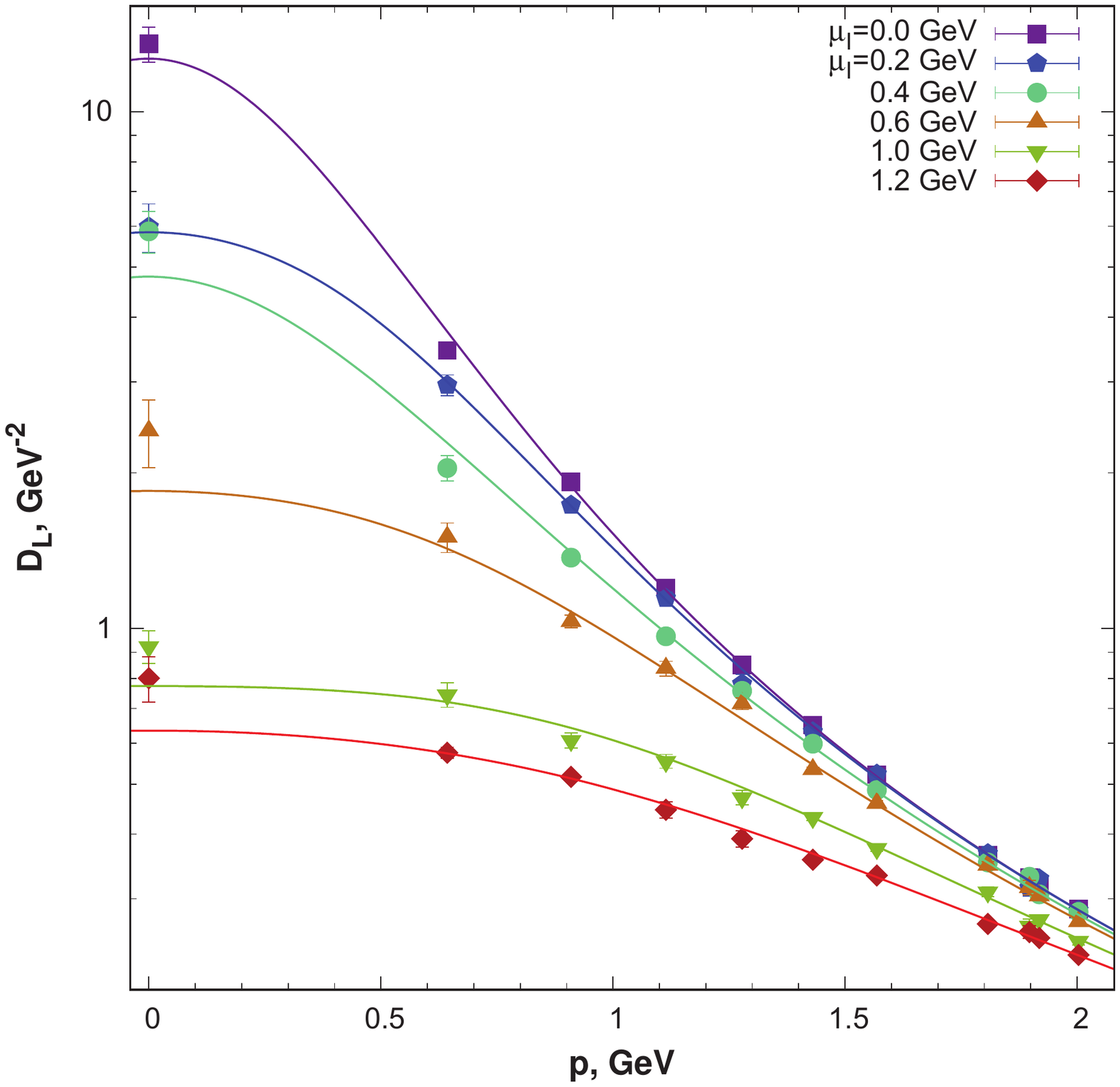}
\includegraphics[width=0.48\textwidth]{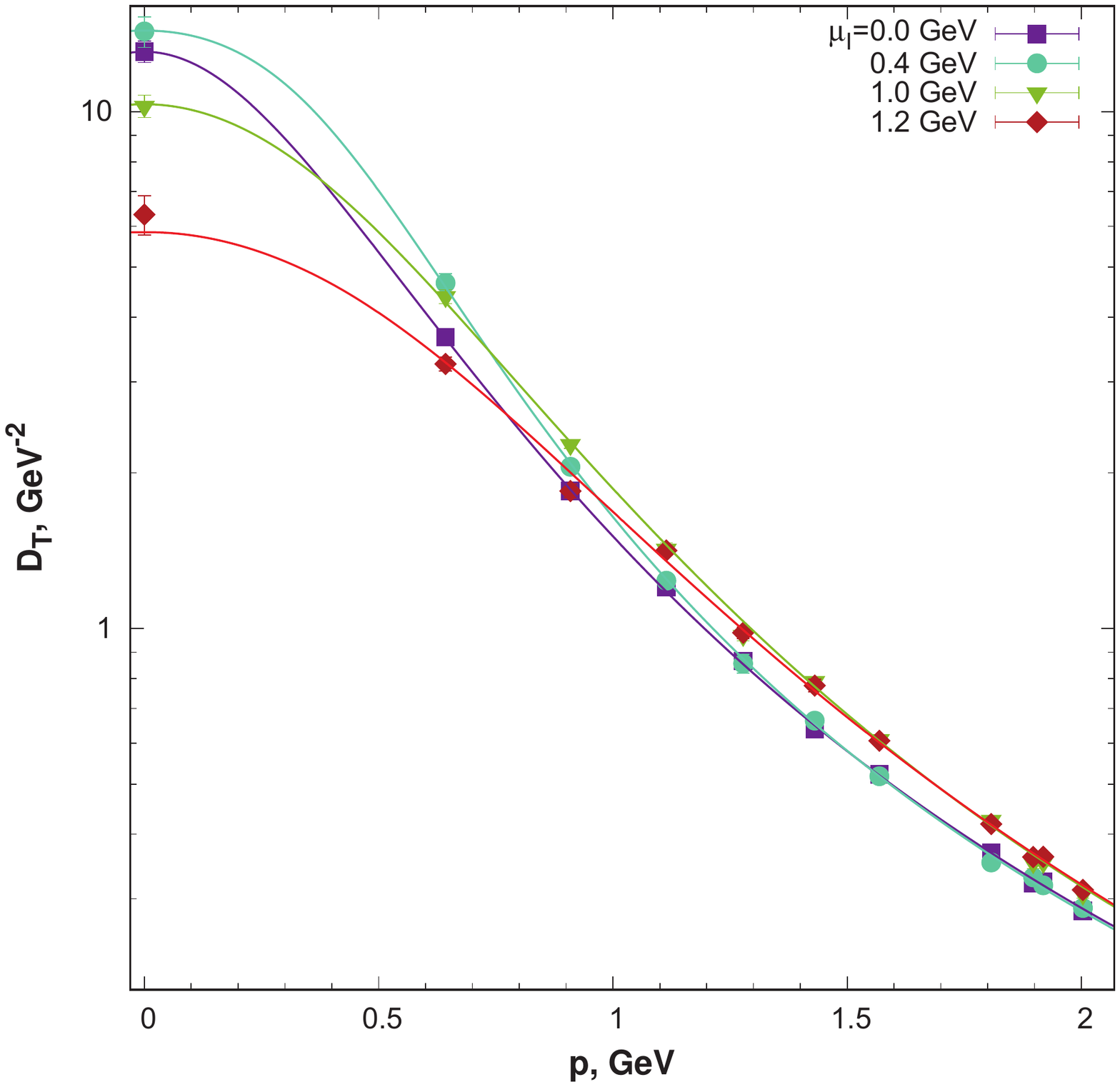}
\vspace*{-25mm}
\caption{The propagators $D_L$ (left) and $D_T$ (right) as functions of $p$ at different values of $\mu_I$. The curves correspond to the formula  (\ref{eq:GS_ff_DOS18}).}
\label{fig:D_vs_p}
\end{figure*}

An appropriate goodness of fit $\ds\Big( {\chi^2/34}=1.66$, 
$p$-value~$\approx 0.009\Big)$ can be achieved for  the momentum range 
$0\leq p < 1.5$~GeV with the following 15-parameter fit function
\bea\label{eq:fit_B_for_mE}
D^{-1}_{L}(p, \mu_I)&\simeq & c_{00} + c_{02} \mu_I^2 
+ c_{04} \mu_I^4 + c_{06} \mu_I^6   \\ \nonumber
&& + c_{20}  p^2 + c_{22} p^2 \mu_I^2 + c_{24}  p^2 \mu_I^4 + c_{26} p^2 \mu_I^6    \\ \nonumber
&& + c_{40} p^4 + c_{42} p^4  \mu_I^2 + c_{44} p^4  \mu_I^4  \\ \nonumber
&& + c_{60} p^6 + c_{62} p^6 \mu_I^2 + c_{64} p^6 \mu_I^4 + c_{80} p^8\;.
\eea
The same fit function provides the best fit quality 
($\ds {\chi^2/27}=1.20$, $p-$value$\approx 0.22$)
when the data at $p=0$ are excluded: $p_{\mbox{min}} < p < 1.5$~GeV,  $0\leq \mu_I \leq 1.2$~GeV.
An increase of $p_{\mbox{cut}}$ to 2~Gev results in some decrease of the $p$-value ($\ds {\chi^2/55}=1.55$, $p-$value$\approx 0.005$).

The square of the chromoelectric screening mass is then determined by the expression
\beq\label{eq:mE_interpolated}
m_E^2(\mu_I) = {c_{00} + c_{02} \mu_I^2 
+ c_{04} \mu_I^4 + c_{06} \mu_I^6  \over  
 c_{20} + c_{22} \mu_I^2 + c_{24} \mu_I^4 + c_{26} \mu_I^6
}
\eeq
The error in this quantity 
which takes into account the correlations
between the fit parameters was calculated 
with the use of the REDUCE computer algebra system. 
The respective corridor of errors 
is shown by the error bars in Fig.\ref{fig:mesq}.

It should be emphasized that the fit (\ref{eq:fit_B_for_mE}) (fit~B) at both values of $p_{cut}$ and the fit (\ref{eq:LOW_MOM_fit_fun}) (fit~A) at $p_{cut}=2.0$~GeV
are stable with respect to an exclusion 
of $p=0$ data. 
This indicates that the dynamics
below $p_{\mbox{min}}$ has only a little effect 
on the screening in the chromoelectric sector.

We show the results of fit~A (\ref{eq:LOW_MOM_fit_fun}) with huge errors 
(for $p_{cut}=1.5$~GeV)
in the left panel of Fig.\ref{fig:mesq} 
in order to illustrate the advantages of
fit~B~(\ref{eq:fit_B_for_mE}).
Some disagreement between the results of the fit~A
and~B is observed, especially in the right panel 
of Fig.\ref{fig:mesq}
at $\mu_I \sim 1.0$~GeV ($p_{cut}=2.0$~GeV).
It should be emphasized that, in performing fit~A, 
the term $\sim p^8$ is not taken into account 
when $p_{cut}=2.0$~GeV
and the terms $\sim p^6$ and $\sim p^8$ are not taken into account when $p_{cut}=1.5$~GeV,
whereas all these terms are included 
in fit function~(\ref{eq:fit_B_for_mE}) in both cases.
The mentioned disagreement indicates that
these terms play a significant role at $p>1$~GeV.
In view of this facts
we conclude that the results 
obtained with fit~B are more reliable
provided that $m_E^2$ is a smooth function of $\mu_I$.
The systematic error of fit~B can be estimated by 
a comparison of the error corridors for the cutoff momenta 
$p_{cut}=1.5$~GeV and $p_{cut}=2.0$~GeV
as well as the error corridor for the case when the zero momentum is eliminated. It is smaller than the statistical error.

One can see in Fig.\ref{fig:mesq} that the chromoelectric screening mass remains approximately constant at $0\leq \mu_I \leq 0.4$~GeV and increases with $\mu_I$ at its greater values.
The dependence of $m_E$ on $\mu_I$ found here is qualitatively similar to its dependence on the temperature
at $T>T_c$ both in pure gluodynamics  and QCD
as was demonstrated by lattice simulations 
in \cite{Bornyakov:2010nc,Silva:2013maa}.
It is also similar to the dependence of $m_E$
on the baryon chemical potential in QC$_2$D
\cite{Bornyakov:2020kyz,Bornyakov:2019jfz}.

Recollecting the dependence
of $m_E$ and $m_M$ on $\mu_q$ and $T$ in $QC_2D$ obtained in \cite{Bornyakov:2021arl}, one is tempted to assume that
the screening in the magnetic sector will increase with $T$
starting from $T\sim 200\div300$~MeV at all values of $\mu_I$,
whereas $m_E$ will show a pronounced increase only at small
values of $\mu_I$. Testing this hypothesis should be the
subject of future studies.

\section{Dependence on the momentum and isospin chemical potential}
\label{sec:GS}

First we consider the momentum dependence of 
the gluon propagators for various values of $\mu_I$ in more detail. The propagators are renormalized according to the MOM 
scheme to satisfy the condition

\beq
D_{L,T}(p=\kappa) = 1/\kappa^2
\eeq           
at $\kappa= 3$~GeV.

In Fig~\ref{fig:D_vs_p} (left) we present the momentum 
dependence for the longitudinal propagator $D_L(p)$ 
for some values of $\mu_I$. In the infrared domain 
it clearly decreases with an increase of $\mu_I$. 
This dependence on $\mu_I$ is similar to the dependence 
of $D_L(p)$ on $T$ at $T>T_c$ in both gluodynamics and 
QCD \cite{Cucchieri:2007ta,Fischer:2010fx,Aouane:2011fv} as well 
as on $\mu_q$ in QC$_2$D \cite{Bornyakov:2020kyz}.

\begin{figure*}[htb]
\hspace*{-16mm}\includegraphics[width=0.7\textwidth]{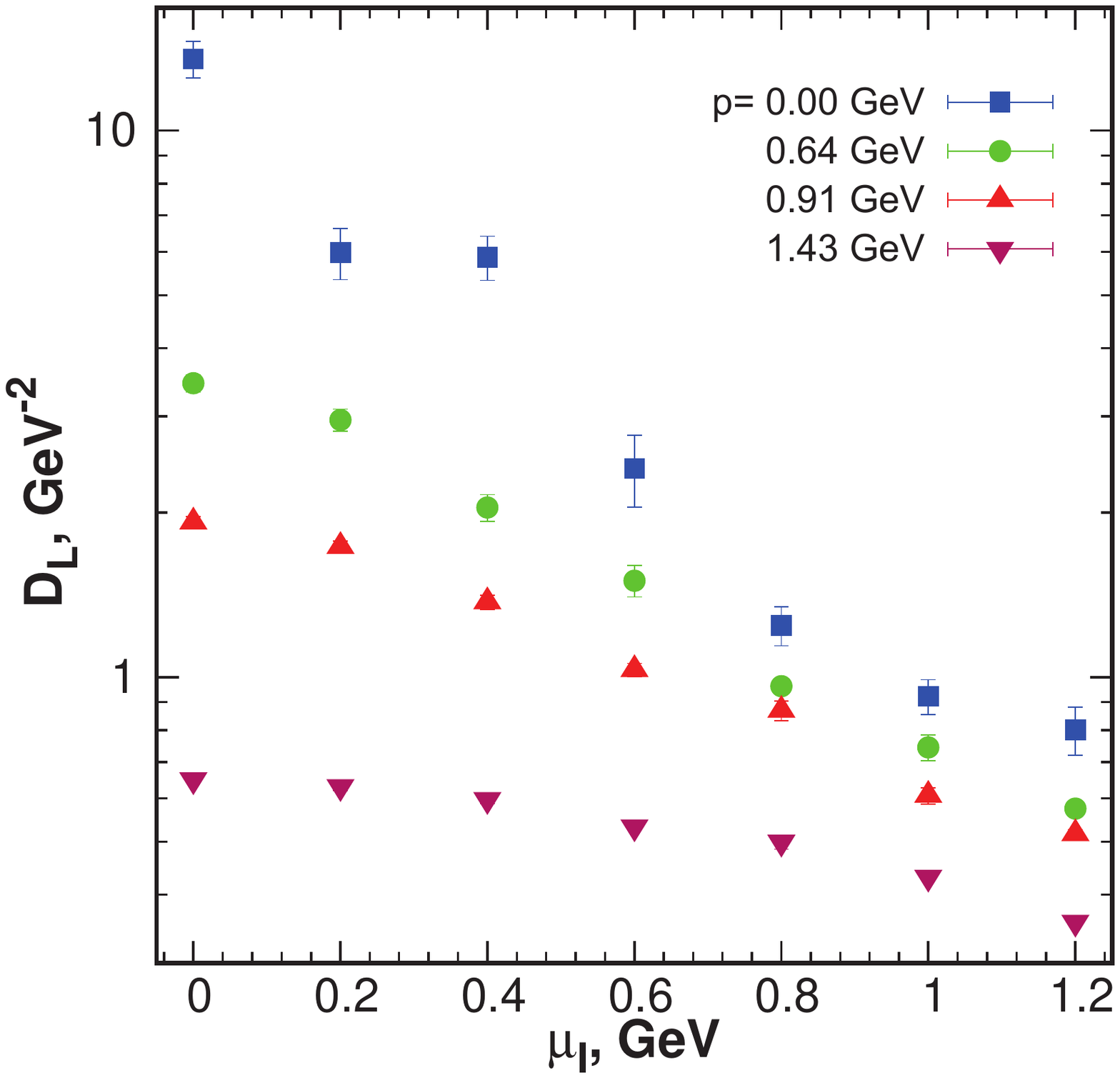}\hspace*{-30mm}\includegraphics[width=0.7\textwidth]{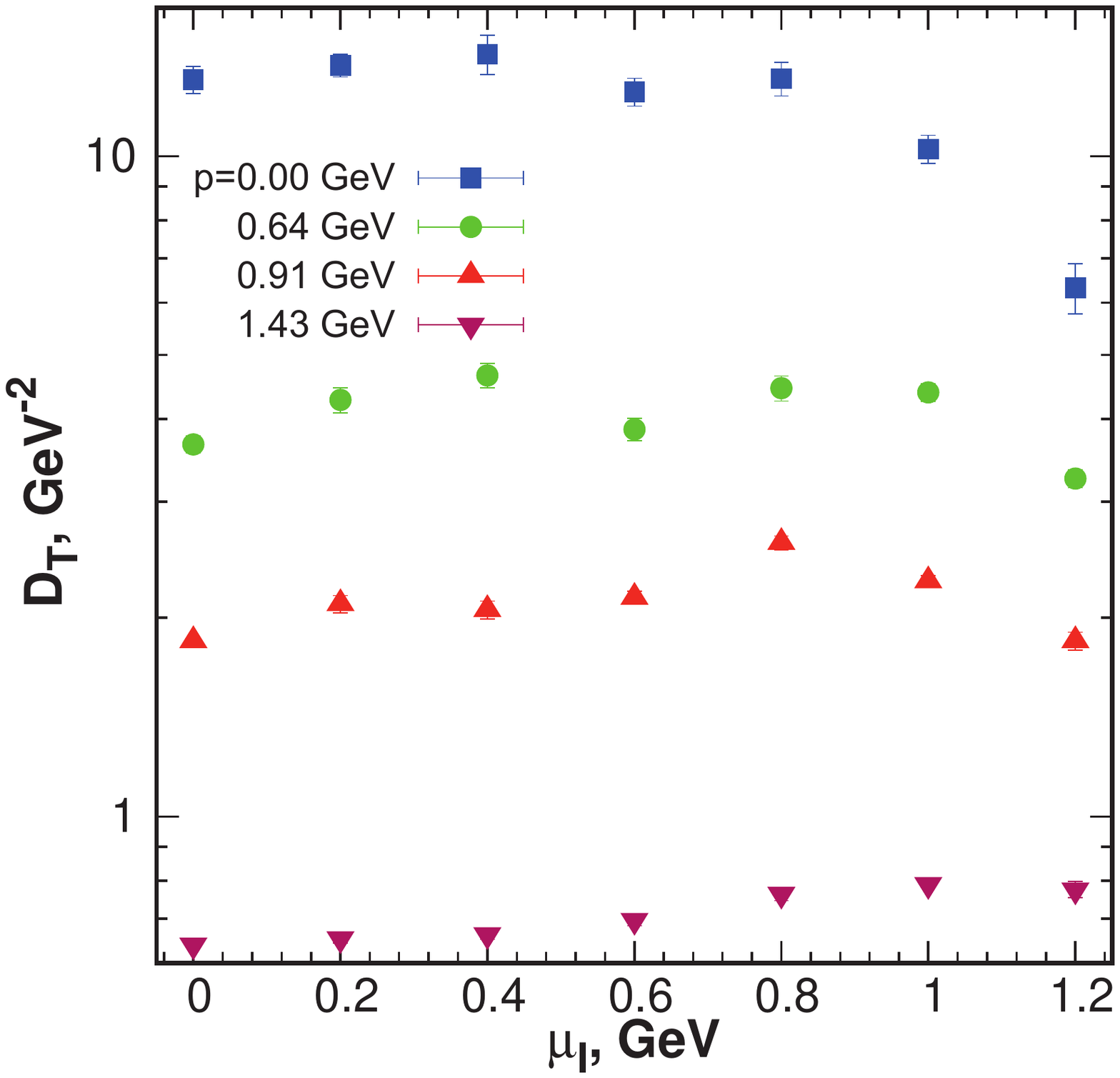}
\vspace*{-15mm}
\caption{The propagators $D_L$ (left) and $D_T$ (right) as functions of $\mu_I$ at different values of $p$. Note logarithm scale on the y axis.}
\label{fig:D_vs_mu}
\end{figure*}

In Fig.~\ref{fig:D_vs_p}(right) the momentum dependence for the transverse propagator $D_T(p)$ for the same values of $\mu_I$ is shown. It is clearly seen that $D_T(p)$ is substantially less sensitive to an increase of $\mu_I$. 
In the deep infrared region  $\mu_I$ dependence of
the transverse propagator differs substantially from that
of the longitudinal propagator. It increases with $\mu_I$ at small $\mu_I$ until it reaches a weakly pronounced peak
(at $\mu_I\simeq 0.4$~GeV for $0\leq p \leq 0.6$~GeV and 
at $\mu_I\simeq 0.8$~GeV at $p=0.9$~GeV, see also Fig.\ref{fig:D_vs_mu}) and then it gradually decreases. At $p>1$~GeV the transverse propagator
increases at intermediate values of $\mu_I: 0.4\leq\mu_I \leq 0.8$~GeV and remains approximately constant at both low and high values of $\mu_I$. A typical dependence of this type is shown in Fig.\ref{fig:D_vs_mu} (right panel) by purple triangles; 
this behavior is also in contrast to that of the longitudinal propagator, which decreases with $\mu_I$ at all momenta as is seen in the left panel of Fig.~ \ref{fig:D_vs_mu} 

It is known that at a finite temperature the propagator  $D_T(p)$ has a clear maximum at the value of momentum increasing with temperature. Our data give no evidence for such maximum at a small momentum, however, we do not rule out its existence.

Now we proceed to an interpolation formula for our data. It was demonstrated  many times
\cite{Dudal:2010tf,Cucchieri:2011ig,Oliveira:2012eh,Dudal:2018cli,Aouane:2011fv}
that the infrared behavior of the gluon propagators
at zero and finite temperature can be well described by the fit function proposed in \cite{Stingl:1994nk} as well as 
the tree level prediction of the Refined Gribov-Zwanziger approach
\cite{Dudal:2008sp},
\beq\label{eq:GS_ff_DOS18}
D(p)= Z \;{M_1^2 +  p^2 \over p^4+ M_2^2 \, p^2 + M_3^4 }\; .
\eeq
This fit function is in fact identical to that used in our 
previous study \cite{Bornyakov:2020kyz},
the relations between the two sets of fit parameters are obvious.
\begin{figure*}[tbh]
\hspace*{-5mm}
\includegraphics[width=9cm]{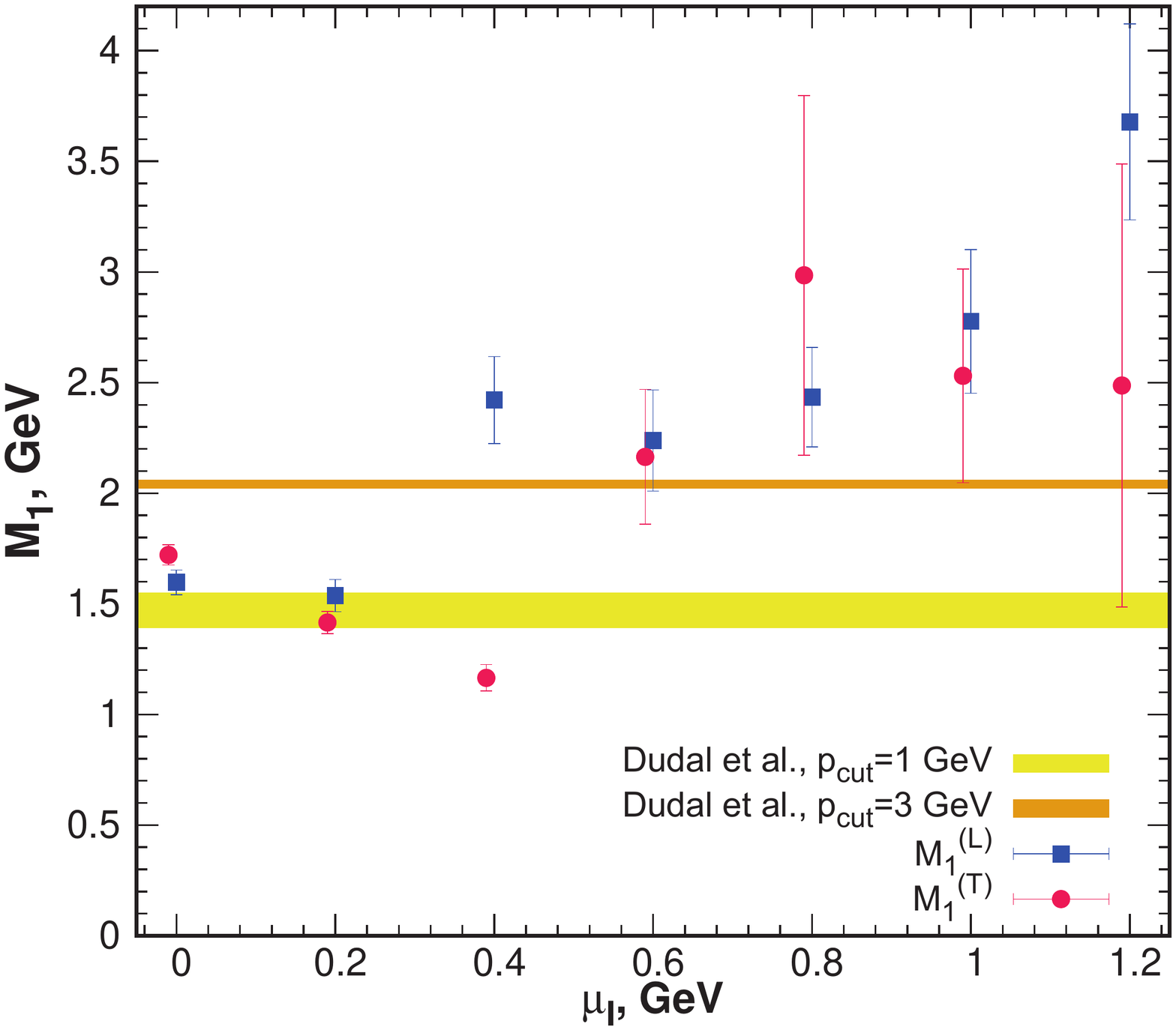}
\hspace*{-15mm}
\includegraphics[width=9cm]{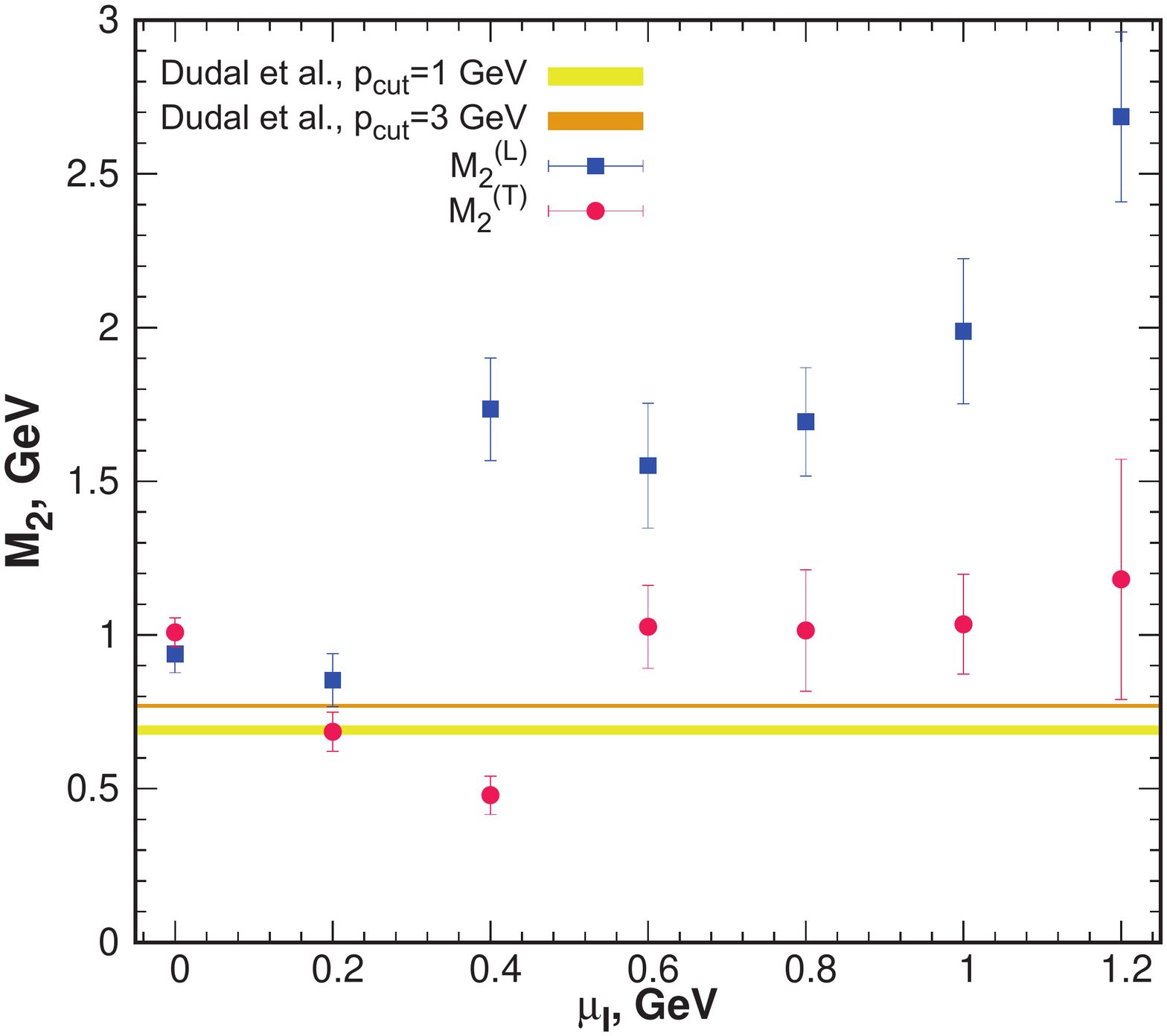}
\vspace*{-6mm}
\caption{Parameters $M_1$ (left) and $M_2$ (right) of the fit (\ref{eq:GS_ff_DOS18}) as functions of $\mu_I$. Yellow strips show
the results obtained in \cite{Dudal:2018cli} with $p_{\mbox{cut}}=1$~GeV, 
brown strips---with $p_{\mbox{cut}}=3$~GeV. }
\label{fig:GS_param_12}
\end{figure*}
\begin{figure}[tbh]
\hspace*{-8mm}\includegraphics[width=9cm]{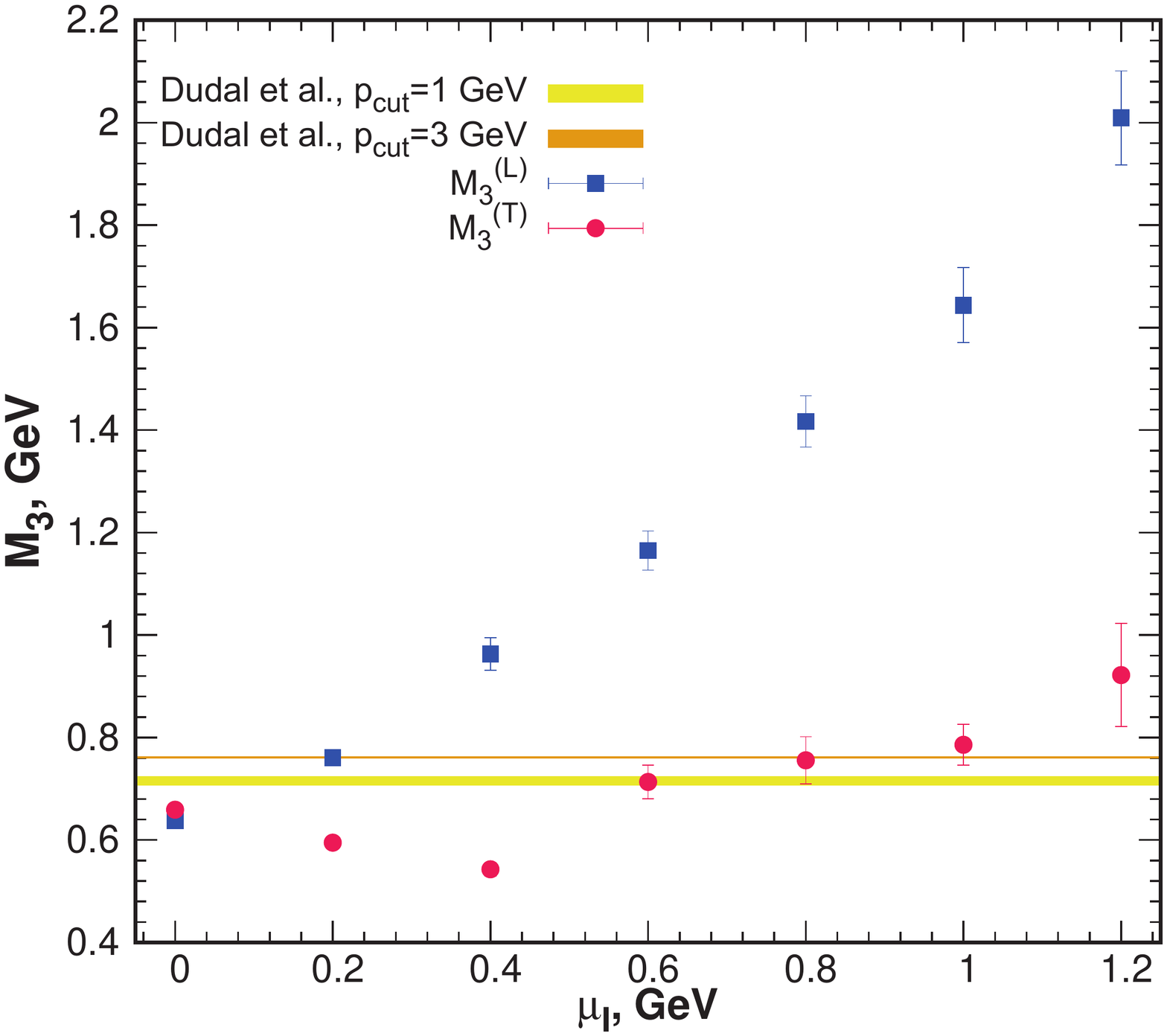}
\vspace*{-6mm}
\caption{Parameter $M_3$ of the GS fit (\ref{eq:GS_ff_DOS18}) as a function of $\mu_I$. Yellow strip shows
the result obtained in \cite{Dudal:2018cli} with $p_{\mbox{cut}}=1$~GeV, 
brown strip---with $p_{\mbox{cut}}=3$~GeV.}
\label{fig:GS_param_3}
\end{figure}

Since the minimal nonzero momentum $p_{\mbox{min}}$ for the lattice under study is rather large our fit over the infrared domain may suffer from finite-volume effects. Nevertheless, we believe that we find qualitatively correct dependence of the propagators on $\mu_I$, at least for $D_L$ at large $\mu_I$.

To lower uncertainties in the fit parameters $M_i$,  we perform fit for the ratio 
\beq
{D(p)\over D(p_0)} \simeq {M_1^2 +  p^2 \over p^4+ M_2^2 \, p^2 + M_3^4 }\; {p_0^4+ M_2^2 \, p_0^2 + M_3^4  \over M_1^2 +  p_0^2}.
\eeq 
where $p_0$ should lie in the fit domain; our choice is $p_0=2$~GeV. 
In so doing, the normalization factors $Z_{L,T}$ are not determined 
and thus should be computed by the formula
\beq
Z={\kappa^4+ M_2^2 \, \kappa^2 + M_3^4 \over \kappa^2 \,(M_1^2 +  \kappa^2) }\;.
\eeq
With this $Z$ factor, formula (\ref{eq:GS_ff_DOS18})
gives the propagator renormalized at $p=\kappa$.

The fit domain in the longitudinal case ranges up to
$p\sub{cut}=5$~GeV; in the transverse case the cutoff momentum
for each value of $\mu_I$ is indicated in
Table~\ref{tab:GST_param_DOS18}.

A few words about the fit stability should be said.
In the longitudinal case, the fit parameters are unaffected
by elimination of zero momentum, in the transverse case 
the fit parameters change significantly at 4 of 7 values 
of $\mu_I$ when zero momentum is excluded.

\begin{figure*}[tbh]
\hspace*{-13mm}\includegraphics[scale=0.4]{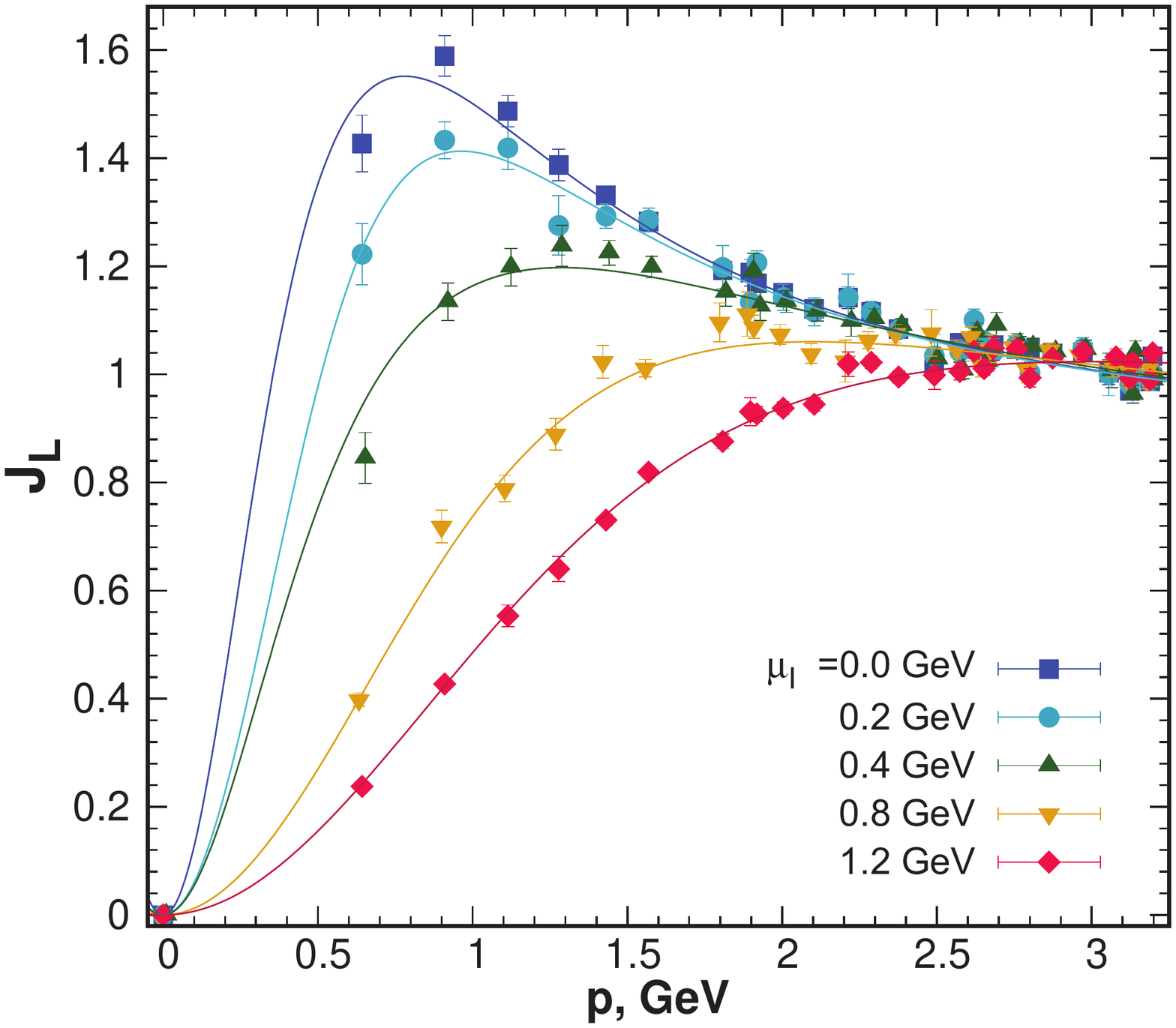}\hspace*{-22mm}\includegraphics[scale=0.4]{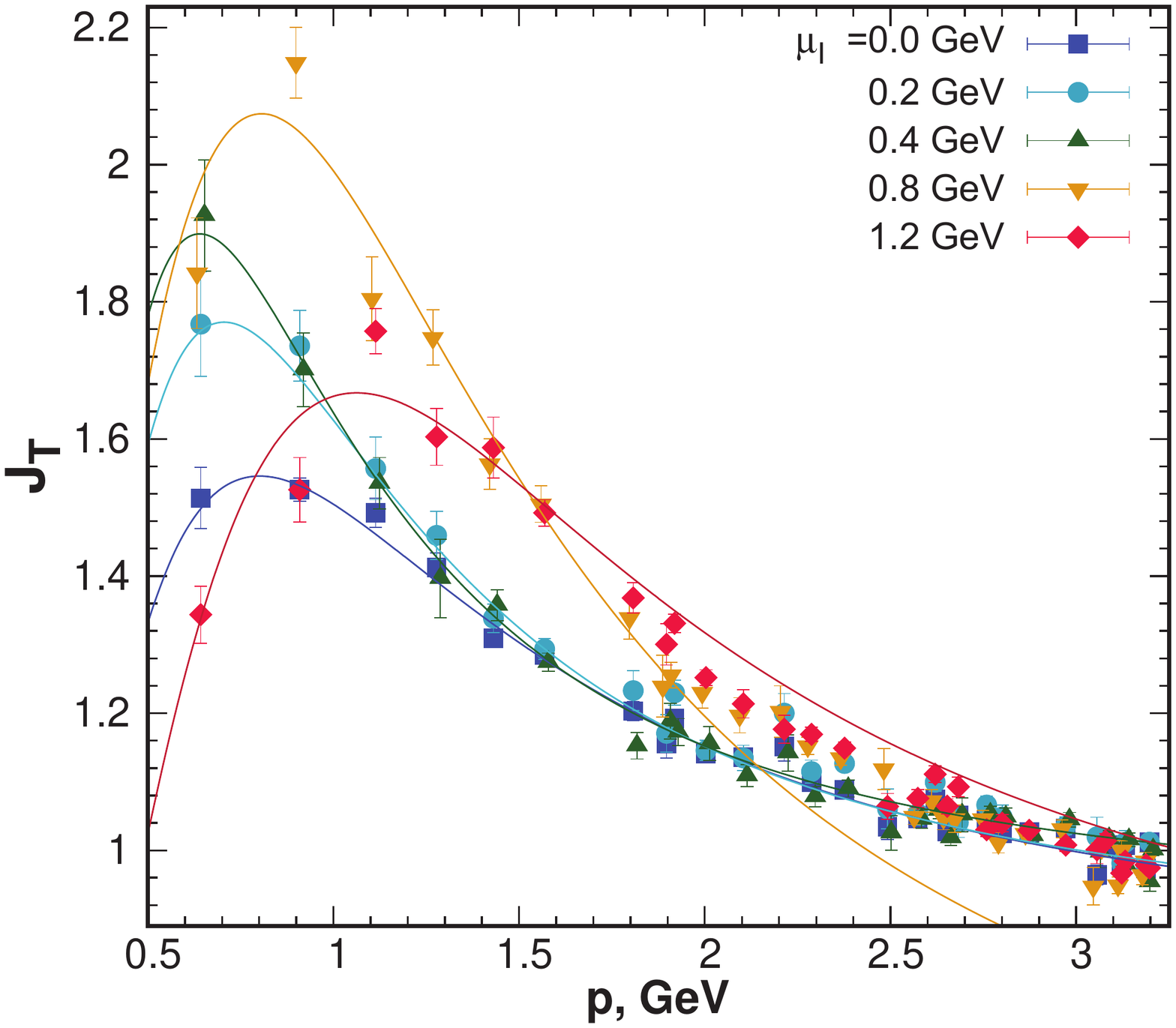}
\vspace*{-12mm}
\caption{The longitudinal (left)and transverse (right) gluon dressing functions at various isospin chemical potentials. 
The curves show results of the fit by (\ref{eq:GS_ff_DOS18}).}
\label{fig:J_vs_p}
\end{figure*}

The fit parameters for $D_L(p)$ depend only weakly 
on the cutoff momentum, as is shown, as an example, 
in Table~\ref{tab:stability_GS_L} for $\mu_I=0$; 
the range $2.2 \mbox{GeV} < p\sub{cut} < 5.0$~GeV is considered 
(at lower value of $p\sub{cut}$ the fit parameters 
are poorly determined). At the other values of 
$\mu_I$ the $p\sub{cut}$-dependence of $M_i$ is also 
insignificant, however, the minimum value of $p\sub{cut}$ 
at which the fit parameters can be determined 
increases with $\mu_I$ and reaches 4.5~GeV at $\mu_I=1.2$~GeV.

The fit parameters and goodness-of-fit for $D_T(p)$ 
depend on $p\sub{cut}$ as follows. At $\mu_I=0.0$ and 0.2~GeV 
we arrive at a reasonable $p$-value for $1.9 < p\sub{cut} < 5.0$~GeV,
$M_i$ depend on $p_{cut}$ only weakly. At $\mu_I=0.4$~GeV,
a reasonable $p$-value for $1.9 < p\sub{cut} < 5.0$~GeV 
also holds, however, $M_i$ depend significantly on $p\sub{cut}$ 
at $p\sub{cut}>3$~GeV. At $0.6\leq\mu_I \leq 1.2$~GeV
we obtain $p$-value~$\geq 0.05$ only at $p\sub{cut}\leq 2.2$~GeV
(this one can see in Fig.~\ref{fig:J_vs_p}(right) 
for the transverse dressing function). 
At $1.9\leq p\sub{cut}\leq 2.2$~GeV $M_i$ 
vary within their errors and at $p\sub{cut}\leq 1.9$~GeV 
they are poorly determined. 

The parameters of this fit are presented in
tables~\ref{tab:GSL_param_DOS18} and~\ref{tab:GST_param_DOS18} 
for the longitudinal and transverse propagators, respectively. 
The parameters $M_1^2$, $M_2^2$, and $M_3^4$ are given in the respective powers of GeV for the sake of comparison with
\cite{Dudal:2018cli}. Results of this fit are also shown in Fig.~\ref{fig:D_vs_p} together with our lattice data.

The dependence of these fit parameters on the 
isospin chemical potential is presented 
in Figs.\ref{fig:GS_param_12} and~\ref{fig:GS_param_3},
where our results are compared with those obtained in
\cite{Dudal:2018cli} in $SU(3)$ gluodynamics at $T=0$
on a large lattice with high statistical precision.

The parameters $M_1^{(L)}$ and $M_2^{(L)}$ have a similar dependence on $\mu_I$: they jump between $\mu_I=0.2$ and $0.4$~GeV and slowly increase at $\mu_I>0.6$~GeV;
the parameter $M_3^{(L)}$ shows approximately linear growth
through the whole range $\mu_I$ under consideration.

\begin{figure*}[tbh]
\hspace*{-17mm}\includegraphics[scale=0.38]{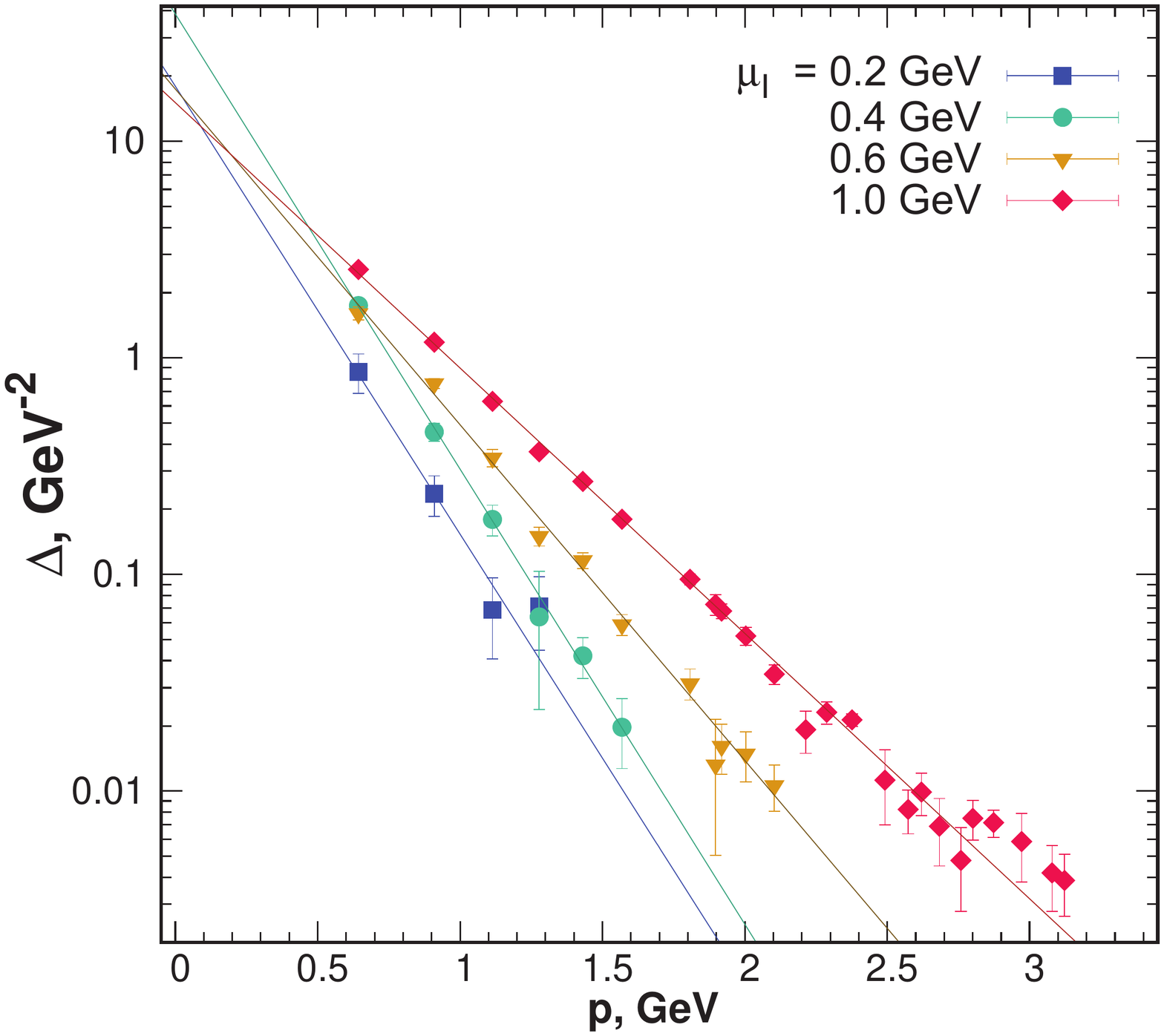}\hspace*{-13mm}\includegraphics[scale=0.38]{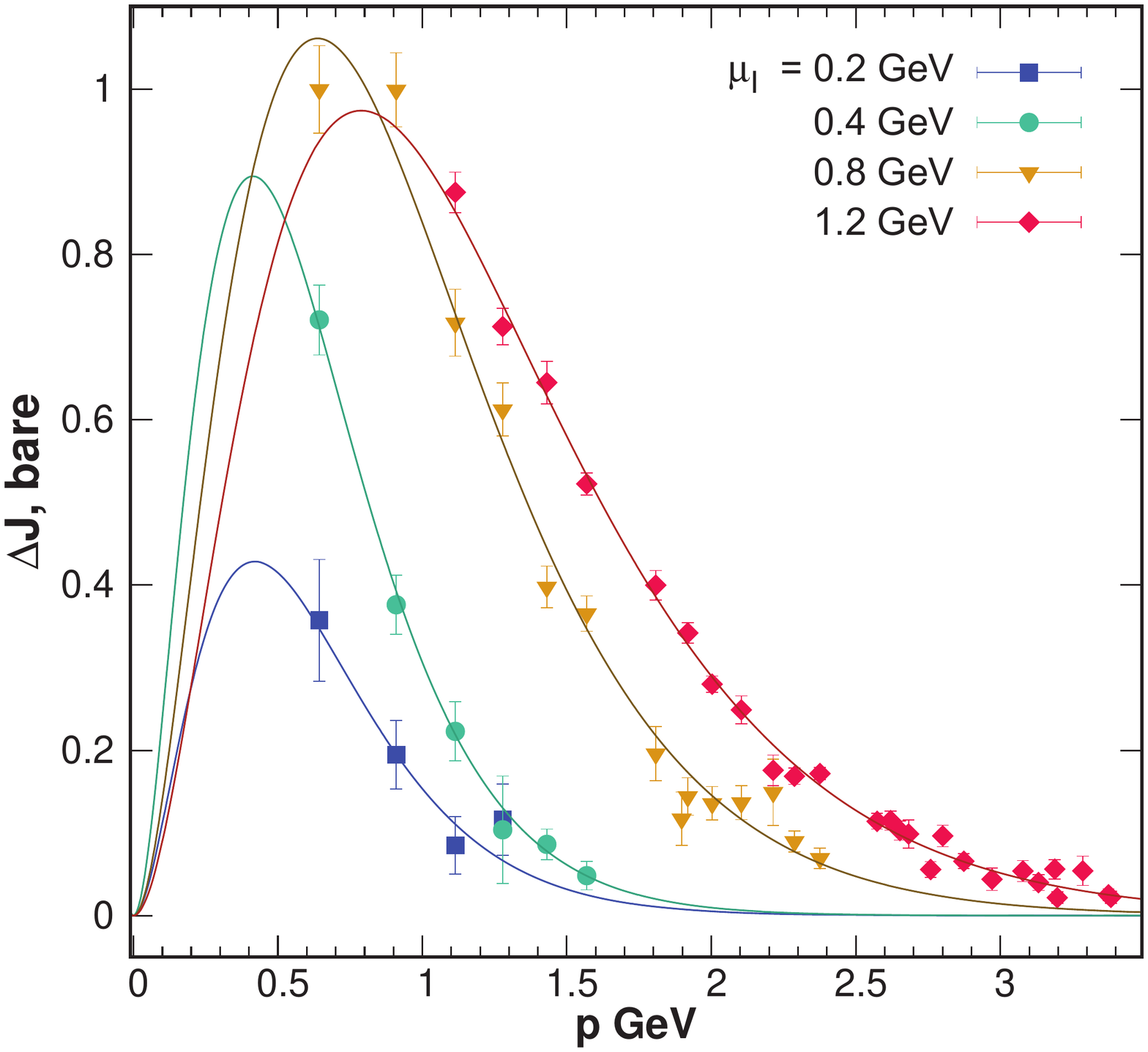}
\vspace*{-12mm}
\caption{The difference between the propagators $D_T-D_L$ (left)
and dressing functions $J_T-J_L$ (right) as a function 
of the momentum. The curves show results of the fit to (\ref{eq:expo_ff}).}
\label{fig:dglp_vs_p}
\end{figure*}

In the case of the transverse propagator all $M_i^{(T)}$ 
approach a shallow minimum at $\mu_I\approx 0.4$~GeV and
remain constant (within error bars) at $\mu_I>0.6$~GeV. A  pronounced difference between $M_i^{(L)}$ and $M_i^{(T)}$
is seen at $\mu_I=0.4$~GeV, this may be caused by the transition  from the normal to the superfluid phase which is expected at $\mu_I\sim 0.2$~GeV.

It is instructive to look also at the respective dressing functions $J_{L,T}(p)$ defined  as
\beq
J_{L,T}(p) = p^2 \, D_{L,T}(p)
\eeq
It is seen in Figure~\ref{fig:J_vs_p} (left) that with increasing  $\mu_I$ the maximum of the longitudinal dressing function goes down and shifts to the right, thus approaching dressing function of a massive scalar particle. We note once more that this dependence on
$\mu_I$ is very similar to dependence on the temperature, see e.g. Ref.~\cite{Fischer:2010fx,Maas:2011se}.

The transverse dressing function increases at $\mu_I < 0.8$~GeV
so that it approaches its peak at $p \sim 0.7\div 0.9$~GeV,
at larger $\mu_I$ the height of the peak decreases with increasing 
$\mu_I$ and the position shifts to $p\sim 1.0\div 1.1$~GeV. 

The difference between the transverse and longitudinal dressing 
functions $\Delta J = J_T-J_L$ is shown in Fig.~\ref{fig:dglp_vs_p}, right panel.
It equals zero at $\mu_I=0$ and then it rapidly increases 
over the range $0<\mu_I< 0.4$~GeV. Then, as $\mu_I$ 
increases from $0.4$ to 1.2~GeV, the height of its peak 
fluctuates over $0.8 < \Delta J < 1.1$,
the width increases and the position shifts 
to higher values of $p$.

It should be emphasized that, at $0<\mu_I< 0.4$~GeV,
when both screening masses remain constant,  
$\Delta J$ shows a rapid change.

\section{Features of $D_L-D_T$ behavior}
\label{section4}

In the previous two sections we demonstrated that the difference between the longitudinal and transverse propagators
increases with an increase of the isospin chemical potential. 
Therewith, they come close to each other at high momenta for a 
fixed value of $\mu_I$. In this section we study 
the rate of the decrease of $\Delta(p)=D_T(p)-D_L(p)$ with increasing momentum and how the picture changes with increasing $\mu_I$.
A similar comparison of these two propagators was made in 
\cite{Bornyakov:2020kyz} in QC$_2$D as well as in \cite{Silva:2013maa} in $SU(3)$ gluodynamics at finite temperatures where the ratio $D_L(p)/D_T(p)$ was considered.

Here we show that, in the theory under study, 
the difference between transverse and longitudinal propagators, $\Delta(p)=D_T(p)-D_L(p)$ at $p_4=0$
shows clear exponential dependence on $p$.

Our numerical results for $\Delta(p)$  are presented 
in Fig.~\ref{fig:dglp_vs_p} (left), together with the fit function
\beq \label{eq:expo_ff}
 \Delta(p) = c \exp (\;-\; \nu \cdot p)\;. 
\eeq
 The exponential decreasing is well established starting from some momentum $p_0$ depending on $\mu_I$.
We found that $0 < p_0 < p\sub{min}$ for $ 0.2 \leq  \mu_I  \leq 1.0$~GeV 
and $p_0 \approx 1.2$~GeV for $\mu_I=1.2$~GeV.
The upper limit of the fit domain $p_0<p<p\sub{cut}$ is 
determined by the requirement that $\Delta(p)$ differs from zero by more than 3 statistical errors (see Table~\ref{tab:expo_fit_param}).

As a check we compare the fit function (\ref{eq:expo_ff}) with the power fit function
\beq\label{eq:pow_ff_2}
\Delta(p) = {C \over p^{v} }
\eeq
motivated by a power-like behavior of both gluon propagators when $p \to \infty$. 

At $0.2\leq \mu_I \leq 0.4$ both fit functions work well
because $\Delta(p)$ does not vanish for only a few momenta. 
At $\mu_I > 0.4$~GeV only the exponential fit function 
(\ref{eq:expo_ff})  works. 
Moreover, we strengthen this conclusion by considering three-parameter power-like fit function as follows:
\beq\label{eq:pow_ff_3}
\Delta(p) \simeq {C \over (p+d)^{v} }\,;
\eeq
it also works only for $0.2\leq \mu_I \leq 0.4$,
the respective $p$-values as well as the fitting parameters in 
formula  (\ref{eq:expo_ff}) are presented in Table~\ref{tab:expo_fit_param}. We also tried several other 
power-like fit functions which also have failed.

Since our data can  be well approximated by the exponential function
and poorly --- by a power-like fit function even with 3 parameters,
we arrive at the conclusion that 
the Gribov-Stingl fit cannot work for $D_L(p)$ and $D_T(p)$ 
simultaneously at large $\mu_I$. Assuming that they do work simultaneously, we obtain
the power-like dependence of the difference $D_T(p)-D_L(p)$,
whereas, as we found, it cannot be fitted by a power-like function.
Of course, the above reasoning is valid provided that the fit domain 
is sufficiently large and statistical precision is sufficiently 
high to discriminate between power-like and exponential behavior.
Given statistical precision as in the case under consideration, 
power-like and exponential behavior are differentiated when
the fit domain is larger than $p\sub{min} \leq p < 2.2$~GeV as one can conclude
from a comparison of Table~\ref{tab:expo_fit_param} 
and Fig.~\ref{fig:dglp_vs_p}. This is the case for 
$\mu_I\geq 0.6$~GeV. Thus our conclusion that 
$\Delta(p;\mu_I)\simeq c \exp (\;-\; \nu \cdot p)$ 
cannot be fitted by a power-like fit function 
at $\mu_I\geq 0.6$~GeV agrees 
with the statement in Section~\ref{sec:GS} 
that the Gribov-Stingl fit (\ref{eq:GS_ff_DOS18}) for $D_T(p)$ 
does not work at these values of $\mu_I$ when $p\sub{cut}>2.2$~GeV,
whereas the same function can be fitted well to $D_L(p)$ 
even at $p\sub{cut}=5$~GeV for all $\mu_I$ under consideration.

We show the curves resulting from our fit (\ref{eq:expo_ff})
to $\Delta(p)$ for some values of $\mu_I$ in Fig.~\ref{fig:dglp_vs_p}
(left panel) as well as the difference between 
the dressing functions
\beq\label{eq:JT-JL}
\Delta J(p) = p^2 \Delta(p) = J_T(p)-J_L(p)
\eeq
(right panel). 

\begin{figure*}[tbh]
\hspace*{-13mm}\includegraphics[scale=0.4]{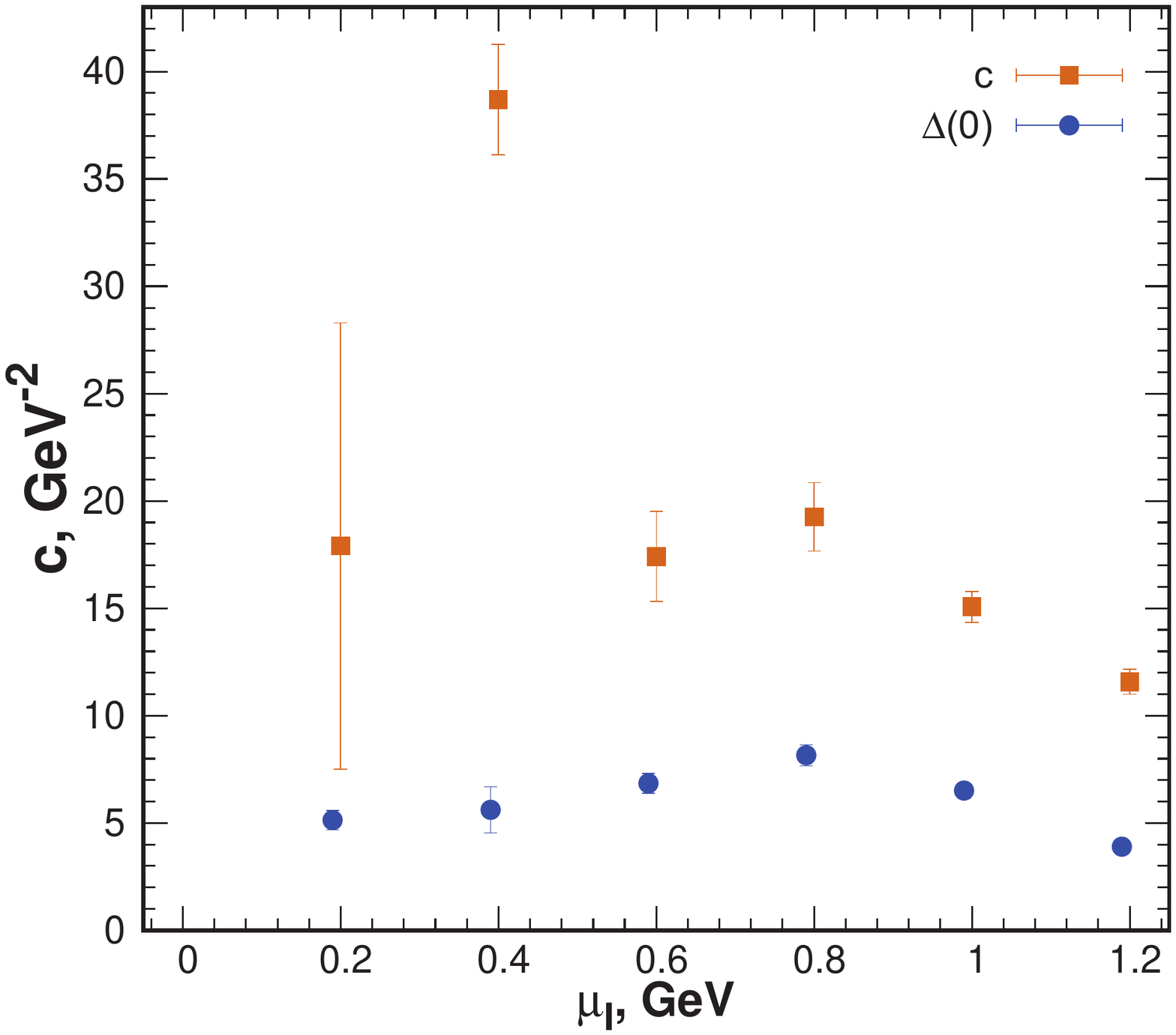}\hspace*{-22mm}\includegraphics[scale=0.4]{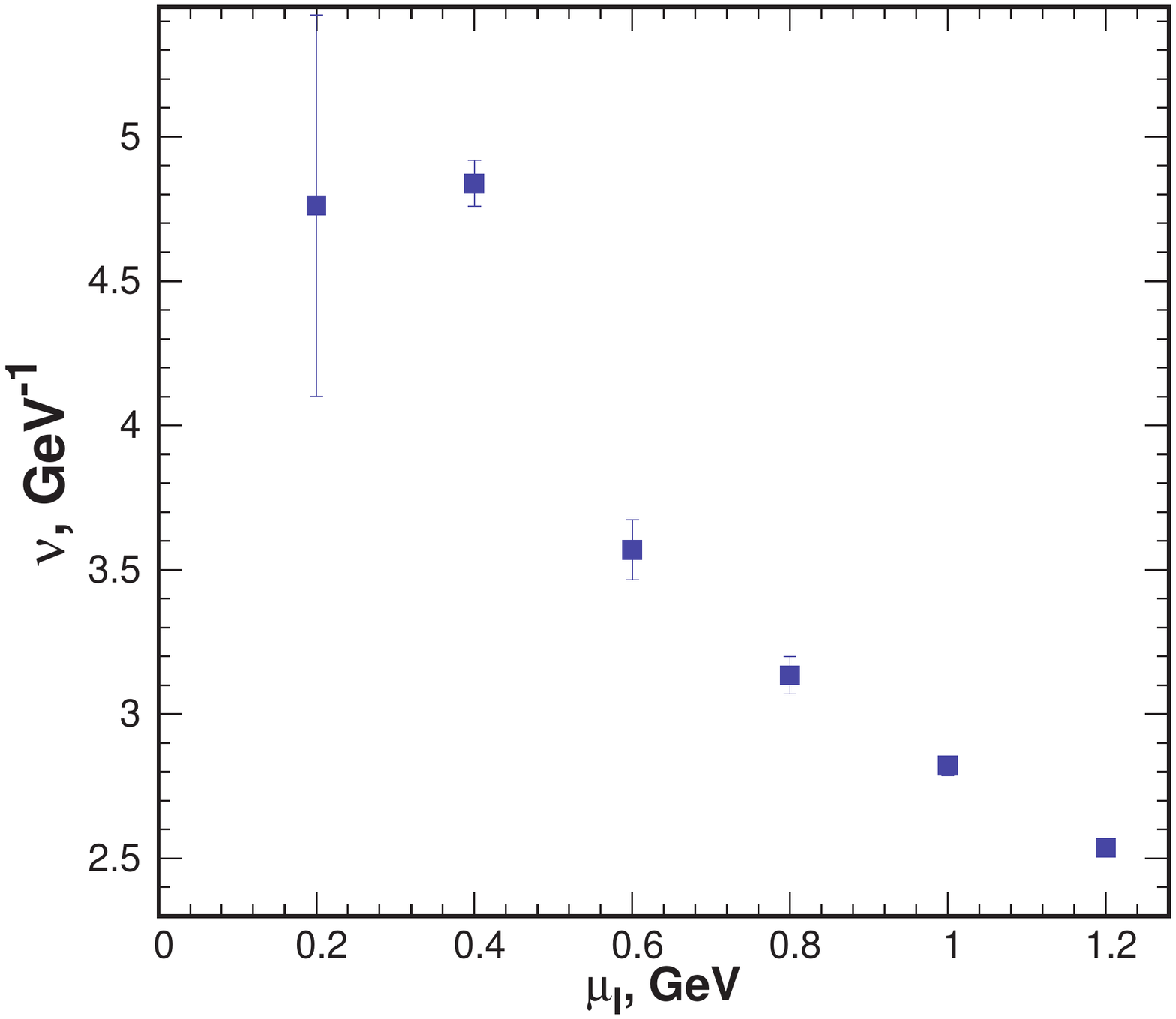}
\vspace*{-12mm}
\caption{The parameters of the fit function (\ref{eq:expo_ff}) as functions of $\mu_I$. In the left panel the parameter $c$
representing the value of $\Delta(0)$ extrapolated by
formula (\ref{eq:expo_ff}) from the fit domain is compared with its actual value. }
\label{fig:c_and_nu_vs_muI}
\end{figure*}

\begin{figure*}[htb]
\includegraphics[width=10cm]{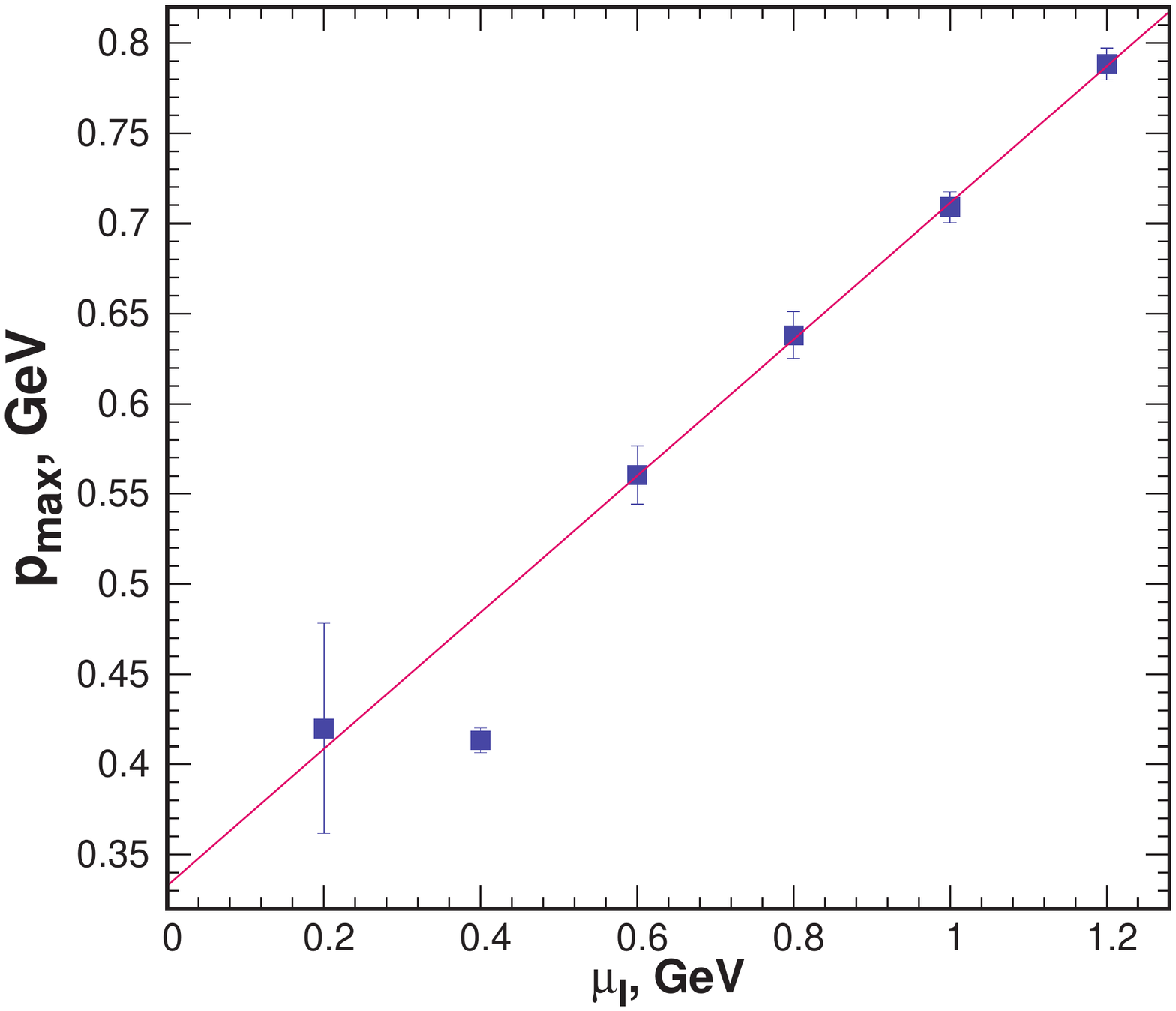}\hspace*{-15mm}\includegraphics[width=10cm]{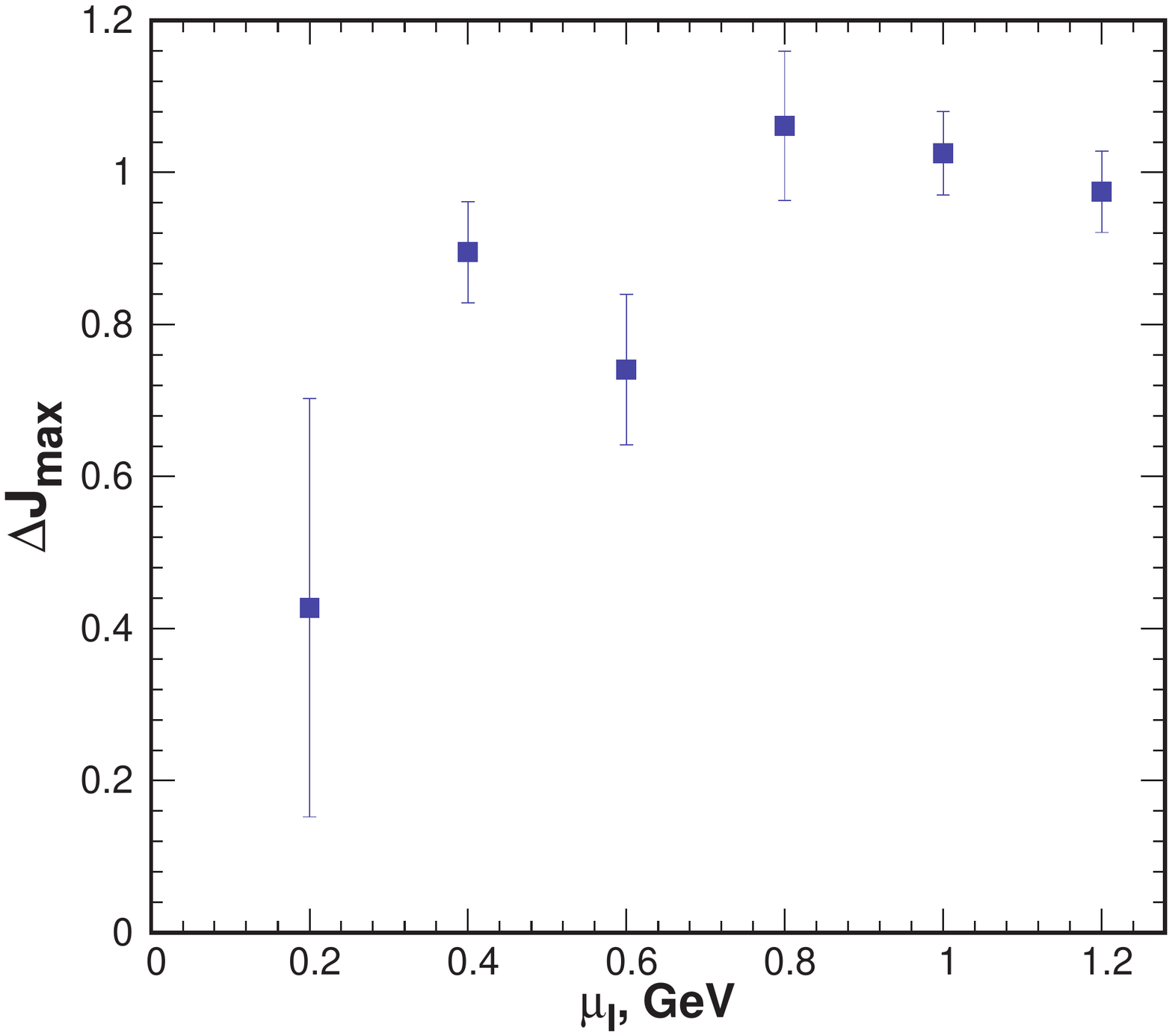}
\caption{Dependence of the position (left panel)
and the height (right panel) of the peak of $\Delta J(p)$ on 
the isospin chemical potential.}
\label{fig:pmax_n_DJmax}
\end{figure*}

The dependence of the parameters $c$ and $\nu$ on the isospin chemical potential is shown in Fig.~\ref{fig:c_and_nu_vs_muI}.
The position and the height of the peak of the difference between the dressing functions (\ref{eq:JT-JL}) is given by the formulas
\beq
p_{max}={2\over \nu}, \qquad \qquad 
\Delta J_{max}={4\over e^2}\;{c^2\over \nu}\;,
\eeq
the respective values are plotted in Fig.~\ref{fig:pmax_n_DJmax}.

It is interesting to notice that, at $\mu_I\leq 0.4$~GeV, 
$\Delta J_{max}$ rapidly increases, whereas 
$p_{max}$ does not change. A completely different situation occurs at $\mu_I \geq 0.4$~GeV: $\Delta J_{max}$ changes only slightly,
whereas $p_{max}$ increases substantially and shows a clear linear dependence on $\mu_I$ at $\mu_I\geq 0.6$~GeV.
That is, the behavior of $p_{max}$ is similar to the behavior
of the electric screening mass (which is associated with the interaction radius), whereas $\Delta J_{max}$ 
(which is associated with the strength of interaction) changes in an opposite manner.

\section{Conclusions}
\label{section5}

We presented first results of the study of the static longitudinal and transverse propagators in the Landau gauge of lattice QCD at nonzero isospin chemical potential.

Our main observations are as follows. We found that the longitudinal propagator $D_L(p)$ is more and more suppressed in the infrared with increasing  $\mu_I$. This is reflected, in particular, in the increase of  the chromoelectric screening mass.
Such dependence of $D_L(p)$ on $\mu_I$ is analogous to its dependence on temperature at $T>T_c$ \cite{Cucchieri:2007ta,Fischer:2010fx,Aouane:2011fv}. It is also similar to dependence of $D_L(p)$ on the quark chemical potential $\mu_q$ in QC$_2$D \cite{Bornyakov:2020kyz}.
We found much  weaker dependence on $\mu_I$ for the transverse propagator $D_T(p)$.
The question of whether the transverse propagator in the infrared domain can be 
adequately described in terms of screening mass
may be the subject of future studies.

We determined the parameters of the
Gribov-Stingl fit function (\ref{eq:GS_ff_DOS18}) 
associated with the Gribov-Zwanziger confinement scenario. 
The parameters for the transverse propagator 
depend only weakly on $\mu_I$ and 
agree well with those in pure gluodynamics 
at $T=0$, the parameters $M_2$ and $M_3$ 
for the longitudinal propagator increase 
with increasing $\mu_I$.
The difference between the parameters of 
the longitudinal and transverse propagators 
approaches a local peak at 
$\mu_I= 0.4$~GeV. This might be caused by the transition 
from the normal to the superfluid phase which is expected at $\mu_I \sim 0.2$~GeV.

We investigated the behavior of the dressing functions, the excess of which over one in the infrared region indicates the presence of  long-range forces. At large values of $\mu_I$ the transverse dressing function is greater than the longitudinal one indicating that the chromoelectric interactions are screened at shorter distances than the chromomagnetic interactions. 

Finally, we  studied the difference $\Delta(p) = D_L(p)-D_T(p)$ and found that it decreases exponentially with the momentum at large $p$ similarly to the case of the QC$_2$D theory
\cite{Bornyakov:2020kyz}.
We also investigated the $\mu_I$ dependence of the parameters of the exponential fit function 
(\ref{eq:expo_ff}) as well as the parameters
$p_{max}$ and $\delta J_{max}$ characterizing 
the behavior of the difference between the longitudinal 
and transverse dressing functions. 
This dependence indicates that
with increasing $\mu_I$ the domain where 
chromomagnetic forces dominate 
over chromoelecric ones extends 
to ever shorter distances.

Our results for the gluon propagators obtained using the first principles approach of the lattice QCD should be definitely useful 
to test other approaches to QCD at finite isospin chemical potential like perturbation theory \cite{Gorda:2021znl}
or functional methods.

{\small {\bf Acknowledgments.}
The authors are grateful to V.V. Braguta and A.Yu. Kotov for very fruitful discussions and to S.A.~Sadovsky for the discussions of statistical analysis. The work was completed due to support of the Russian Foundation for Basic Research via grant 18-32-20172~mol-a-ved and grant 18-02-40130 mega.  
The research was carried out using the Central Linux Cluster of the NRC "Kurchatov Institute" - IHEP,  the Linux Cluster of the NRC "Kurchatov Institute" - ITEP (Moscow) and 
the equipment of the shared research facilities of HPC computing resources at Lomonosov Moscow State University. In addition, we used computer resources of the federal
collective usage center Complex for Simulation and Data Processing for
Mega-science Facilities at NRC “Kurchatov Institute”, http://ckp.nrcki.ru/.
}

\vspace*{182mm}
\appendix \section{Fit results}
\label{sec:appx}
\nopagebreak[4]
\begin{table*}[hbt]
\caption{Values of the parameters of the Gribov-Stingl fit
(\ref{eq:GS_ff_DOS18}) for the longitudinal propagator $D_L(p^2)$
at various $\mu_I$. The cutoff momentum $p\sub{cut}=5$~GeV for all
values of $\mu_I$}
\label{tab:GSL_param_DOS18}
\begin{tabular}{lllllll} \hline
$\mu_I,$~GeV~ & ~$\mu_I a$~ &~Z~&$M_1^2$ &$M_2^2$ &$M_3^4$     &$p$-value  \\ \hline
0.0  & 0.000 & 0.864(11) & 2.55(18)  &  0.88(11) &  0.165(14) &    0.26    \\
0.2  & 0.070 & 0.865(10) & 2.36(23)  &  0.73(15) &  0.334(26) &    0.56    \\
0.4  & 0.139 & 0.826(10) & 5.86(95)  &  3.01(58) &  0.86(11)  &    0.28    \\
0.6  & 0.209 & 0.838(10) & 5.0(1.0)  &  2.41(63) &  1.84(24)  &    0.18    \\
0.8  & 0.279 & 0.840(10) & 5.9(1.1)  &  2.87(60) &  4.03(57)  &    0.89    \\
1.0  & 0.348 & 0.840(11) & 7.7(1.8)  &  3.95(94) &  7.3(1.3)  &    0.15    \\
1.2  & 0.418 & 0.819(17) & 13.5(3.3) &  7.2(1.5) &  16.3(3.0) &    0.63    \\
\hline
\end{tabular}
\end{table*}

\begin{table*}[hhh]
\caption{Values of the parameters of the Gribov-Stingl fit (\ref{eq:GS_ff_DOS18}) for the transverse propagator $D_T(p^2)$.
}
\label{tab:GST_param_DOS18}
\begin{tabular}{lllllll} \hline
$\mu_I$,~GeV~&~Z~&$M_1^2$~ &$M_2^2$&~~$M_3^4$~ &$p-$~ & $p\sub{cut},$ \\ 
         &   &           &             &           & value &  GeV          \\ \hline
0.0  & 0.837(8) & 2.96(16)  &  1.02(10) &  0.189(11) &  0.47 & 5.0 \\
0.2  & 0.861(8) & 2.01(14)  &  0.47(9)  &  0.125(9)  &  0.04 & 5.0 \\
0.4  & 0.914(9) & 1.36(14)  &  0.23(6)  &  0.087(8)  &  0.91 & 3.0 \\
0.6  & 0.687(7) & 4.7(1.3)  &  1.05(28) &  0.259(47) &  0.68 & 2.2 \\
0.8  & 0.473(5) & 8.9(4.9)  &  1.03(40) &  0.326(79) &  0.08 & 2.2 \\
1.0  & 0.619(7) & 6.4(2.4)  &  1.07(34) &  0.381(78) &  0.18 & 2.2 \\
1.2  &0.724(11) & 6.2(5.0)  &  1.40(92) &  0.72(31)  &  0.05 & 1.9 \\
\hline
\end{tabular}
\end{table*}

\begin{table*}[htb]
\caption{ Dependence of the parameters $M_i$, 
determined by a fit of the formula (\ref{eq:GS_ff_DOS18}) to $D_L(p)$
over the range $0\leq p < p\sub{cut}$ on the cutoff momentum 
$p\sub{cut}$ at $\mu_I=0$.}
\label{tab:stability_GS_L}
\begin{tabular}{lllll} \hline
$p\sub{cut},$~GeV~&$M_1^2$~ &$M_2^2$~ &$M_3^4$~ &$p-$value~ \\ \hline
2.2  &  3.59(83)  &  1.20(27) &  0.214(36) &  0.92  \\
2.6  &  2.67(50)  &  0.90(20) &  0.172(26) &  0.74  \\
3.0  &  2.04(23)  &  0.67(11) &  0.139(14) &  0.86  \\
3.5  &  2.10(18)  &  0.69(10) &  0.143(13) &  0.70  \\
4.0  &  2.26(18)  &  0.75(11) &  0.151(13) &  0.39  \\
4.5  &  2.43(19)  &  0.82(12) &  0.159(14) &  0.20  \\
5.0  &  2.55(18)  &  0.88(12) &  0.165(14) &  0.25  \\
\hline
\end{tabular}
\end{table*}

\begin{table*}[bbh]
\caption{Parameters $c$ and $\nu$ as well as $\Delta J_{max}$ 
determined by fitting formula (\ref{eq:expo_ff}) to our data for $\Delta(p)$. In the last two columns the values of the goodness-of-fit parameter corresponding to (\ref{eq:expo_ff}) and (\ref{eq:pow_ff_3})
are compared.}
\label{tab:expo_fit_param}
\begin{tabular}{lllllll} \hline
$\!\mu_I\!$ &$\!p\sub{cut},$ &$c,$     &$\nu,$    &$\Delta J_{max}$& $p$-value           & $p$-value\\ 
GeV$\!\!\!$ &        GeV     &GeV$^{-2}$&GeV$^{-1}$&                &eq.(\ref{eq:expo_ff})&eq.(\ref{eq:pow_ff_3}) \\ \hline
0.2  & 1.3 & 17.9(10.4) &  4.76(66)  &  0.52(21)  &  0.39  & 0.57 \\
0.4  & 1.6 & 38.7(5.0)  &  4.839(80) &  0.91(13)  &  0.96  & 0.43 \\
0.6  & 2.2 & 17.4(2.1)  &  3.57(10)  &  0.738(76) &  0.086 & 0.002 \\
0.8  & 2.5 & 19.3(1.6)  &  3.135(64) &  1.066(65) &  0.056 & 0.0004 \\
1.0  & 3.1 & 15.07(72)  &  2.821(54) &  1.014(64) &  0.039 & $10^{-8}$ \\
1.2  & 3.4 & 11.58(58)  &  2.537(28) &  0.978(49) &  0.067 & 0 \\
\hline
\end{tabular}
\end{table*}

\end{document}